\documentclass[preprintnumbers,article,amsmath,amssymb,floatfix,10pt,prd,onecolumn,
superscriptaddress,nofootinbib]{revtex4}
\usepackage{bbm}
\usepackage{amsfonts}
\usepackage{mathrsfs}
\usepackage{latexsym}

\usepackage{epsfig}
\usepackage{graphicx}
\usepackage{epstopdf}
\usepackage{placeins}
\usepackage{float}

\usepackage{amssymb}
\usepackage{amsmath}

\usepackage{dcolumn}
\usepackage{bm}
\usepackage{color}
\usepackage{comment}
\usepackage{xcolor}

\begin{document}

\title{\bf Bardeen Compact Stars in Modified \textbf{$f(R)$} Gravity with Conformal Motion}

%
%
%

%
%
%

\author{M. Farasat Shamir}
\email{farasat.shamir@nu.edu.pk}\affiliation{National University of Computer and
Emerging Sciences,\\ Lahore Campus, Pakistan.}

\author{Aisha Rashid}
\email{aisharashid987@gmail.com}\affiliation{National University of Computer and
Emerging Sciences,\\ Lahore Campus, Pakistan.}


\begin{abstract}
The main emphasis of this paper is to find the viable solutions of Einstein Maxwell fields equations of compact star in context of modified $f(R)$ theory of gravity. Two different models of modified $f(R)$ gravity are considered. In particular, we choose isotropic matter distribution and Bardeen's model for compact star to find the boundary conditions as an exterior space-time geometry. We use the conformal Killing geometry to compute the metric potentials. We discuss the behavior of energy density and pressure distribution for both models. Moreover, we analyze different physical properties such as behavior of energy density and pressure, equilibrium conditions, equation of state parameters, causality conditions and adiabatic index. It is noticed that both $f(R)$ gravity models are suitable and provides viable results with Bardeen geometry.

\end{abstract}

\maketitle

\section{Introduction}

The exploration of compact star is the most striking area for the researchers in the last few years. The usual endpoint of stellar evolution is the formation of a compact star when outward radiation pressure from the nuclear fusion in its interior can no longer resist the gravitational forces. The quarks star, white dwarfs, brown dwarfs, neutrons and black holes are the class of compact star. Some regular black holes exits which are singularity free. In 1968 Bardeen provided Bardeen black hole \cite{Bar} which is the resolution of black hole by using spherically symmetry space time. Frenando and Corrrea \cite{Fer} have analyzed quasi-normal modes of Bardeen star by using scalar perturbations. Which provided the desirable results in different directions. Sharif and Javed \cite{shj} have presented quantum corrections of Bardeen Black hole. A study on electromagnetic and gravitational stability has been discussed by Moreno and Sarbach \cite{msr}. Eiroa and Sendra \cite{eir} have studied the gravitational lensing of consistent black hole. In presence of charge the values of mass, radius and redshifts may be affected. Shamir and Mustafa \cite{sha3} have developed the new family of charged anisotropic compact star using Bardeen's analysis as an exterior space time. Kumar et al \cite{kum} have analyzed the exact thermodynamics quantities of Bardeen black hole. Gedala et al \cite{ged} discovered two new exact and analytic solution of Einstein Maxwell field equation in the context of Bardeen black hole also satisfying embedding conditions. Mustafa et al \cite{mus} presented new interior solution using Kermarker conditions in light of Bardeen stellar structure.

Modified theories of gravity prolong the form of general relativity through different methods using different field equation and cosmological effects. They play vital role in modern cosmology, providing base for the current understating of physical phenomena of the universe. A number of  theories have been introduced by different researchers. For example, theories like $f(R)$ \cite{sha2,carr,cap,sha5,nor,bert,odi}, $f(R,G)$ \cite{ata,nor2}, $f(R,T)$ \cite{har,das} and $f(R,\phi,X)$ \cite{sha4} have been developed by combining curvature scalars and their derivatives. Einstein's general relativity has been generalized by a kind of gravitational theory which depends on random function $f$ of Ricci scalar ($R$). As a result of introducing random function, there may be liberty to explain the speedy  expansion and establishment of the universe structure without accumulation unknown forms of dark energy or black objects. Bunchdahl \cite{bun} presented $f(R)$ modified theory of gravity in 1970. Nojiri and Odintsov \cite{noj} discussed general properties of $f(R)$ gravity along with popular models and their applications for the combined explanation of inflation with dark energy. Cognola et al \cite{cog} has presented a class of modified $f(R)$ exponential models along with stability of presented models. The discussion of compact stars in modified theories of gravity can be interesting task

In non-perturbation $f(R)$ gravity \cite{asta22}, quark star structures with equation of state (EoS) are explored. The mass-radius relationship for the $f(R)$ model is found  along with scalar-tensor gravity equivalent description. However, its contribution  to gravitational mass is small for distant observers.The maximum limit of neutron star mass increases for particular EoS, and hence such EoS can characterize realistic stellar structures \cite{asta44,capo222}. The maximum mass of neutron stars (NS) and the minimum mass of black holes are essential concerns that have been addressed by the authors in \cite{asta11} by estimating the maximum baryonic mass of NSs in the framework of $f(R)$ gravity using multiple EoS models. They have shown that the minimum  limit for the mass of astrophysical black holes is  $M \sim 2.5$ to $3M_{\odot}$ in context of $GW190814$ gravitational wave \cite{asta33}.

Symmetries has been used to find the relation between the geometry and matter using Einstein field equations. Killing vectors are the examples of such symmetries. Herrera's \cite{herr} contributions are a staple in the adoption of categories for a one group of conformal motion. Conformal symmetries are useful in evolving mathematical related issues for some solutions in compact star. Herrera et al \cite{herr,herr1,herr2} have investigated the solution avowing one parameter group of conformal motion for anisotropic matter. He also showed that for the group of conformal motion, the equation of state is Inimitably define through the Einstein equations. Herrera et al \cite{herr2} found some analytical solutions of Einstein-Maxwell filed equations for perfect fluid. They matched all solutions to RN metric which possess positive energy density larger than stresses within the sphere. Shamir \cite{Sha} has discussed the solution of Einstein-Maxwell equations by choosing Bardeen model admitting conformal motion. He investigated the equilibrium parameters by using TOV equation and stability of stellar structure in presence of charged. Esculpi and Aloma \cite{esc} developed two new families of solution of charged anisotropic compact star maintaining one parameter group of conformal motion. They considered tangential pressure to be proportional to radial pressure. Mak and Harko \cite{mak} found the solution with conformal symmetries of static, spherically gravitational field equation in the vacuum in the brane. They have taken conformal factor as parameter. Shamir and Malik \cite{sha2} developed the solution of charged compact star using modified $f(R)$ gravity. They used the matching condition of spherically symmetric space-time with Bardeen geometry to find the exterior solution. Shamir and Mustafa \cite{sha3}, in this paper they found new family  of charged anisotropic compact star solution using conformal killing symmetries.

This paragraph includes all the framework of current research paper. Section II includes the field equation of modified theory of gravity $f(R)$, which have been developed along with the formulation of effective energy density and effective pressure. In section III, the conformal killing vectors are defined and the metric coefficients have been developed by using the definition of conformal killing vectors. In section IV, details of both presented models of modified $f(R)$ gravity have been discussed. In section  V, the values of constants have been defined by using matching condition of Bardeen's exterior solution. In section VI, we have presented the physical analysis  of our models along with graphical representation in detail. In the last section conclusion and comparison have been discussed for both models.

 \section{EINSTEIN-MAXWELL Field EQUATION }
 The Hilbert action in presence of charge is as follows
 \begin{equation}\label{1}
 S=\int d^{4}x\sqrt{-g}\large[\dfrac{1}{16 \pi G} f(R)+L_{e}+L_{m}\large]
 \end{equation}
 In above action  $f$ is the function of Ricci scalar $R$,  $L_{m}$ represents the Lagrangian, $g$ is the determinant of line element $g_{\mu\nu}$ and $L_{e}$ is the lagrangian for electromagnetic field is given by
 \begin{equation}\label{2}
 L_{e}=\dfrac{1}{16\pi}F_{\alpha\beta}F_{\xi\psi}g^{\alpha\xi}g^{\beta\psi},
 \end{equation}
 an electromagnetic field tensor $F_{\alpha\beta}$ with electromagnetic four potential vector $A_{\alpha}$ is given as
 \begin{equation}\label{3}
 F^{\alpha\beta}=A_{\alpha,\beta}-A_{\beta,\alpha},
 \end{equation}
 Einstein Maxwell field equations are as follows
 \begin{equation}\label{4}
 F^{\alpha\beta}_{;\beta}=-4 \pi j^{\alpha}.
 \end{equation}
 In above Einstein$-$Maxwell field equation $j^{\alpha}=\sigma\nu^{\alpha}$ is the electromagnetic four current vector with $\sigma$ the charge density. In this research we have taken spherically symmetric space-time of the form
\begin{equation}\label{5}
ds^2=-(e^\lambda dr^2 +r^2 d\theta^2 +r^2\sin^2\theta d\phi^2)+e^\chi dt^2 ,
\end{equation}
With occurrence of charge the stress energy momentum tensor is given by
\begin{equation}\label{6}
T_{\chi\xi}=(\rho+p)u_{\chi}u_{\xi}-pg_{\chi\xi}-\dfrac{1}{4\pi}\large(g_{\xi\beta}F^{\alpha\beta}F_{\xi\alpha}-\dfrac{1}{4}g_{\chi\xi}F_{\alpha\psi}F^{\alpha\psi})\large.
\end{equation}
 The energy density, radial pressure and tangential pressure are represented by $\rho$, $p_{r}$ and $p_{t}$ respectively. The velocity four vectors are satisfying the following conditions

 \begin{equation}\label{7}
 u^{\chi}u_{\chi}=1 ,\nu^{\chi}\nu_{\chi}=-1
 \end{equation}
 There is only one component of Einstein-Maxwell equation that exists is $F^{01}$. By working on Eq.($\ref{3}$)and Eq.($\ref{4}$) we get the following
 \begin{equation}\label{8}
 F^{01}=\dfrac{q}{r^{2}}e^{-\frac{\chi+\lambda}{2}},
 \end{equation}
 where $q$ is the total charge which is a function of $r$.
 \begin{equation}\label{9}
 q=4\pi \int4 \sigma \omega^{2} e^{-\frac{\lambda}{2}}d\omega.
 \end{equation}
We obtain the electric field intensity $E$ as
\begin{equation}\label{10}
E=-F^{01}F_{01}=\dfrac{q}{r^{2}}.
\end{equation}
The modified field equation of $f(R)$ gravity is as follows
\begin{equation}\label{11}
{G^{eff}}_{\chi\xi}=\kappa{T^{eff}}_{\chi\xi}
\end{equation}
where
\begin{equation}\label{12}
T^{eff}_{\chi\xi}=\dfrac{1}{f_{R}}(\kappa T_{\chi\xi}+E_{\chi\xi}+\dfrac{1}{2} g_{\chi\xi}f(R)+\triangledown_{\chi}\triangledown_{\xi}f_{R}-\dfrac{1}{2}R f_{R}g_{\chi\xi}-g_{\chi \xi} \square f_{R})
\end{equation}

All above information have been used to determine the field equations of the following form
\begin{equation}\label{13}
\rho^{eff}+\dfrac{E^{2}}{f_{R}}=\dfrac{1}{f_{R}} \Big( \kappa \rho+\dfrac{f-Rf_{R}}{2}+\dfrac{1}{e^{\lambda}} \Big[f^{''}_{R}+(\frac{\lambda^{'}}{2}+\frac{2}{r})f^{'}_{R} \Big] \Big ),~~~~~~~~
\end{equation}
\begin{equation}\label{14}
p^{eff}-\dfrac{E^{2}}{f_{R}}=\dfrac{1}{f_{R}} \Big( \kappa p+\dfrac{f-Rf_{R}}{2}-\dfrac{f^{'}_{R}}{e^{\lambda}}(\frac{\chi^{'}}{2}+\lambda^{'}+\frac{2}{r})\Big),~~~~~~~~~~~~~~~~
\end{equation}

\begin{equation}\label{16}
\sigma=\dfrac{e^{-\frac{\lambda}{2}}}{4 \pi r^{2}}(r^{2} E)^{'}.
\end{equation}

\section{Killing Vectors and Conformal Motion}
Conformal killing vector can provide the best results as it gives a deeper understanding of space geometry. The work which have been done previously on these equation (discussed in section I) shows that conformal symmetry in the manifold is useful technique to establish the exact solution of Einstein Maxwell equation. The conformal killing vector are given by the following equation
\begin{equation}\label{0}
\pounds_{\xi}g_{a b}=\phi g_{a b}
\end{equation}

where $a$, $b=0,1,2,3$ showing four dimensional space-time and $\pounds_{\xi}g_{a b}$ is the Lie derivative of a metric tensor with respect to a vector field $\xi$. Now using Eq.($\ref{0}$) along with the space time ($\ref{5}$) we acquire the following conformal killing equations
\begin{equation}\label{17}
\xi^{0}=L,~~~~~~\xi^{1}\chi^{1}=\phi,~~~~~\xi^{1}=\dfrac{\phi r}{2},~~~~~~\xi^{1}\lambda^{'}+2\xi^{1}_{,1}=\phi.
\end{equation}
By solving Eqs.($\ref{17}$) simultaneously we get the following metric potentials
\begin{equation}\label{18}
e^{\chi}=X^{2}r^{2},~~~~~e^{\lambda}=(\dfrac{Y}{\phi})^{2},~~~~~\xi^{i}=W \delta^{i}_{4}+(\frac{r\phi}{2})\delta^{i}_{1},
\end{equation}
where $X$, $Y$ and $W$ are arbitrary constants. we suppose $\phi$ as a non-zero linear function of $r$ i.e.  $\phi=X+Zr$, where $Z$ is a random constant. We also assume the electric charge of the form given as
\begin{equation}\label{19}
E^{2}=\dfrac{g k r}{g r+1}
\end{equation}
where $g$ and $k$ are the arbitrary constants.

\section{Realistic Models of $f(R)$ theory gravity}
In the current paper the analysis has been done by using two different realistic and viable models of $f(R)$ gravity. The details are given in the following sections.
\subsection{Model I}
First we take the starobinsky like model \cite{star2} given by
\begin{equation}\label{162}
f(R)=R+mR^{2},
\end{equation}
where we take $m$ as constant which shows exponential growth of cosmic expansion.It is of great significance as all graph depends on parameter m. \\
By using Eqs.($\ref{13}$) to($\ref{16}$), Eq.($\ref{18}$) and Eq.($\ref{19}$) we attain the physical quantities given as
\begin{eqnarray}
&& \rho^{eff}=\dfrac{1}{r^{2}(1+g r)Y^{2}(r^{2}Y^{2}+4 m(-Y^{2}+3(X+r Z)(X+2r Z)))}[ -2 m (1+gr)(-3 X^{4}-2 X^{2} Y^{2}+Y^{4}+ \nonumber\\
&& 6 r X(3 X^{2}-Y^{2})Z+r^{2}(75 X^{2}-4 Y^{2})Z^{2}+90 r^{3}X Z^{3}+36 r^{4} Z^{4})+r^{4} Z^{4}(K \rho +g r(-k +k \rho))],\label{a1}\\
&& p^{eff}=\frac{1}{r^2 (1+g r) Y^2 (r^2 Y^2+4 m (-Y^2+3 (X + r Z)(X + 2 r Z)))}[g k r^2 Y^5+ K p r^4(1+ g r)Y^4+ 2 m (1+ g r) \nonumber\\
&& (45 X^4 -18 X^2 Y^2 + Y^4 + 2 r X(78 X^2 - 17 Y^2) Z +r^2 (201 X^2 - 16 Y^2) Z^2+ 126 r^3 X Z^3+36 r^4 Z^4)]\label{b1}
\end{eqnarray}

\subsection{Model II}
In present research we choose second model of $f(R)$ gravity for the analysis of compact star with conformal motion is logarithmic corrected $R^2$ model as discussed in \cite{sha5} and  \cite{odi} given as
\begin{equation}\label{191}
f(R)= R+ \alpha R^{2}+\beta R^{2}ln(\beta R)
\end{equation}
where $\alpha$ and $\beta$ are arbitrary constants.
By using  Eqs.($\ref{13}$) to($\ref{16}$), Eq.($\ref{18}$) and Eq.($\ref{19}$) we get the physical quantities given as

\begin{eqnarray}
&& \rho^{eff}=\dfrac{1}{a_1 r^2 (1+g r) Y^2(r^2 Y^2+a_3(2 \alpha +\beta)+2 a_3 \beta \log(\frac{a_3 \beta}{r^2 Y^2}))}(-a_1 g k r^5 Y^4-2 a_1^3 g r \alpha +4 a_1 a_2 X^2 \alpha +4 a_1 a_2 g r X^2 \alpha+\nonumber\\
&& 12 a_1 a_2 r X Z \alpha +12a_1 a_2g r^2 X Z \alpha -36 a_1 r X^3 Z \alpha - 36 a_1 g r^2 X^3 Z \alpha + 8 a_1 a_2 r^2 Z^2 \alpha+ 8 a_1 a_2 r^3 Z^2 \alpha -72 a_1 r^2 X^2 Z^2 \alpha -\nonumber\\
&&72 a_1 g r^3 X^2 Z^2 \alpha -36 a_1 r^3 X Z^3 \alpha -36 a_1 g r^4 X Z^3 \alpha -2 a_1^3 \beta -2 a_1^3 g r \beta +6 a_1 a_2 X^2 \beta +4 a_2^2 X^2 \beta+6 a_1 a_2 g r X^2 \beta + \nonumber\\
&&4 a_2^2 g r X^2 \beta +18 a_1 a_2 r X Z \beta +8a_2^2 r X Z \beta +18 a_1 a_2 g r^2 X Z \beta +8 a_2^2 g r^2 X Z \beta -54 a_1 r X^3 Z \beta -54 a_1 g r^2 X^3 Z \beta \nonumber\\
&&+12 a_1 a_2 r^2 Z^2 \beta + 4 a_2^2 r^2 Z^2 \beta +12 a_1 a_2 g r^3 Z^2 \beta +4 a_2^2 g r^3 Z^2 \beta-108 a_1 r^2 X^2 Z^2 \beta -108 a_1 g r^3 X^2 Z^2 \beta -54 a_1 r^3 X Z^3 \beta \nonumber\\
&& -54 a_1 g r^4 X Z^3 \beta + a_1 \kappa r^4 Y^4 \rho +a_1 g \kappa r^5 Y^4 \rho -a_1^3 \beta \log(4)-2 a_1^3 (1+g r)\beta \log(\frac{a_3 \beta}{r^2 Y^2}))\\
&& p^{eff}=\dfrac{1}{r^2 (1+g r) Y^2(r^2 Y^2+a_3(2 \alpha +\beta)+2 a_3 \beta \log(\frac{a_3 \beta}{r^2 Y^2}))}(\kappa p_r r^4 Y^4+ g k r^5 Y64 +g \kappa p_r r^5 Y^4-2 a_1^2 \alpha +2 a_1 a_3 \alpha -  \nonumber\\
&&2 a_1^2 g r \alpha +2 a_1 a_3 g r \alpha +12 a_2 X^2 \alpha  +12 a_2 g r X^2 \alpha +16 a_2 r X Z \alpha +16 a_2 g r^2 X Z \alpha +4 a_2 r^2 Z^2 \alpha +4 a_2 g r^3 Z^2 \alpha + a_1 a_3 \beta \nonumber\\
&& +a_1 a-3 g r \beta +18 a_2 X^2 \beta +18 a_2 g r X^2 \beta +24 a_2 r X Z \beta +24 a_2 g r^2 X Z \beta + 6 a_2 r^2 Z^2 \beta + 6 a_2 g r^3 Z^2 \beta -2 a_1^2(1+g r) \beta  \log(\frac{2 a_1 \beta}{r^2 Y^2})\nonumber\\
&&+ 2(1+g r)(a_1 a_3 +6 a_2 X^2+ 8 a_2 r X Z +2 a_2 r^2 Z^2 )\beta  \log(\frac{a_3 \beta}{r^2 Y^2})) )
\end{eqnarray}

where
\begin{eqnarray*}
a_1 &&= 3 X^2-Y^2+9r X Z + 6 X^2 Z^2, a_2= 6X^2-2 Y^2+9 r X Z, a_3= 2(-Y^2+3(X+rZ)(X+2rZ)).
\end{eqnarray*}

\section{Matching with Bardeen Model}
In current section, for finding the solution we match spherically symmetric space time with Bardeen exterior space time, which is defined as
\begin{equation}\label{20}
ds^{2}=j(r)^{-1}d r^{2}+r^{2}d \theta^{2}+r^{2}sin^{2}\theta d^{2}-j(r)dt^{2},
\end{equation}
where
\begin{equation}\label{21}
j(r)=1-\dfrac{2Mr^{2}}{(q^{2}+r^{2})^{\frac{3}{2}}},
\end{equation}
Where $M$ is the mass of usual astrophysical formation, the asymptotic behavior of Bardeen star is given by
\begin{equation}\label{22}
j(r)=1-\dfrac{2M}{r}+\dfrac{3Mq^{2}}{r^{3}}+O(\dfrac{1}{r^{5}}).
\end{equation}

In above equation, the term having fraction of $r^{3}$ makes this model different from RN solution. We have taken $j(r)=1-\dfrac{2M}{r}+\dfrac{3Mq^{2}}{r^{3}}$ for current analysis.  We are analyzing the interior and exterior geometries for physical aspects.Wwe use the following continuity conditions for the  metric potentials on the boundary $(r=R_{b})$ and the solution obtained from conformal killing vector Eq.(\ref{18}). We have the matching conditions,
\begin{equation}\label{23}
1-\dfrac{2M}{r}+\dfrac{3Mq^{2}}{r^{3}}=X^{2}r^{2}
\end{equation}
\begin{equation}\label{24}
\Big(1-\dfrac{2M}{r}+\dfrac{3Mq^{2}}{r^{3}}\Big)^{-1}=(\frac{Y^{2}}{(X+Z r)^{2}})
\end{equation}
\begin{equation}\label{25}
\frac{\partial g^{-}_{tt}}{\partial r} = \frac{\partial g^{+}_{tt}}{\partial r}
\end{equation}

Where $(+)$ and $(-)$ shows the exterior and interior solution respectively. By using Eqs.(\ref{23})-(\ref{25}) we get,
\begin{equation}\label{26}
X=- \sqrt{\dfrac{R_b(1+g R_b)-2M(1+g R_b)+3m g k R_b^3}{R_b^3+g R_b^4}},
\end{equation}

\begin{equation}\label{27}
Y=-\dfrac{\sqrt{-2M+ R_b-2gM R_b+g R_b^2+3 g k M R_b^3}-Z R_b^{\frac{5}{2}} \sqrt{1+g R_b}}{-R_b \sqrt{-2 M+ R_b -2 g M R_b+ g R_b^2+3 g k M R_b^3}},
\end{equation}

\begin{equation}\label{28}
Z=-\dfrac{2 \sqrt{1+g R_b}(R_b + g R_b^2-2M-2Mg R_b+3Mgk R_b^3)^{\frac{3}{2}}}{R_b^{\frac{5}{2}}(2 R_b(1+g R_b)^2-6M-12Mg R_b-6Mg^2 R_b^2+3Mg^2 k R_b^3)}
\end{equation}
These parameters plays significant role for defining the behavior of charged star. To match interior space-time we discuss the appropriate  boundary conditions.

\section{Physical Analysis}
In current section, For physical validity of stellar structure, we illustrated the graphical representation of energy density, pressure distribution, EoS parameters, energy conditions, equilibrium conditions, mass$-$radius ratio and adiabatic parameter. For model-I we fix the constant $k=0.8$ and the detail of the rest of constants are given in table \ref{tab:1}. For model-II we fix the constant $\alpha$ and table \ref{tab:2} shows the details. The analysis have been done by choosing massive stars with radius not less than $9$ and maximum masses must be with in $M \sim 2.5$ to $3M_{\odot}$ as described in \cite{asta11,asta33,Sha}.

\begin{table}[ht]
\caption{Values of unknown constants for $k=0.8$.}
\centering
\begin{tabular}{|p{2.3cm}|p{1.8cm}| p{1.2cm}| p{1.2cm}| p{2.3cm}| p{2.3cm}| p{2.3cm}|}
\hline\hline
$M(M_{\odot})$ & $R_b$ & $m$ & $g$ & $X$ & $Y$ & $Z$ \\
\hline
$2.97979661$ & $9.12904$ \cite{Sha} & 0.10 & $4$ & $-0.205752$ & $-0.142473$ & $-0.00873881$  \\
\hline
$2.97644068$ & $9.14507$ \cite{Sha} & 0.20 & $4.2$ & $-0.203263$ & $-0.142542$ & $-0.00864716$  \\
\hline
$2.96298998$ & $9.20902$ \cite{Sha} & 0.30 & $3.9$ & $-0.203996$ & $-0.14254$ & $-0.00867793$  \\
\hline
$2.94349831$ & $9.30119$ \cite{Sha} & 0.40 & $4$ & $-0.202295$ & $-0.142611$  & $-0.00861998$ \\
\hline
$2.93093559$ & $9.36021$ \cite{Sha} & 0.50 & $4$ & $-0.2033218$ & $-0.142602$ & $-0.00865751$  \\
\hline
\end{tabular}
\label{tab:1}
\end{table}

\begin{table}[ht]
\caption{Values of unknown constants for $g=0.3$ and $\alpha=0.2$.}
\centering
\begin{tabular}{|p{2.3cm}|p{1.8cm}| p{1.2cm}| p{1.2cm}| p{2.3cm}| p{2.3cm}| p{2.3cm}|}
\hline\hline
$M(M_{\odot})$ & $R_b$ & $k$ & $\beta$ & $X$ & $Y$ & $Z$ \\
\hline
$2.97979661$ & $9.12904$  & 0.6 & $0.2$ & $-0.770323$ & $-0.138778$ & $-0.0306343$  \\
\hline
$2.97644068$ & $9.14507$  & 0.7 & $0.4$ & $-0.83146$ & $-0.138755$ & $-0.0330464$  \\
\hline
$2.96298998$ & $9.20902$ & 0.8 & $0.5$ & $-0.883538$ & $-0.137379$ & $-0.0347578$  \\
\hline
$2.94349831$ & $9.30119$  & 0.9 & $0.5$ & $-0.943703$ & $-0.141243$  & $-0.0381605$ \\
\hline
$2.93093559$ & $9.36021$  & 0.9 & $0.6$ & $-0.939609$ & $-0.140396$ & $-0.0377677$  \\
\hline
\end{tabular}
\label{tab:2}
\end{table}

\subsection{Energy Density and Pressure}
\begin{figure}\center
\begin{tabular}{cccc}
\epsfig{file=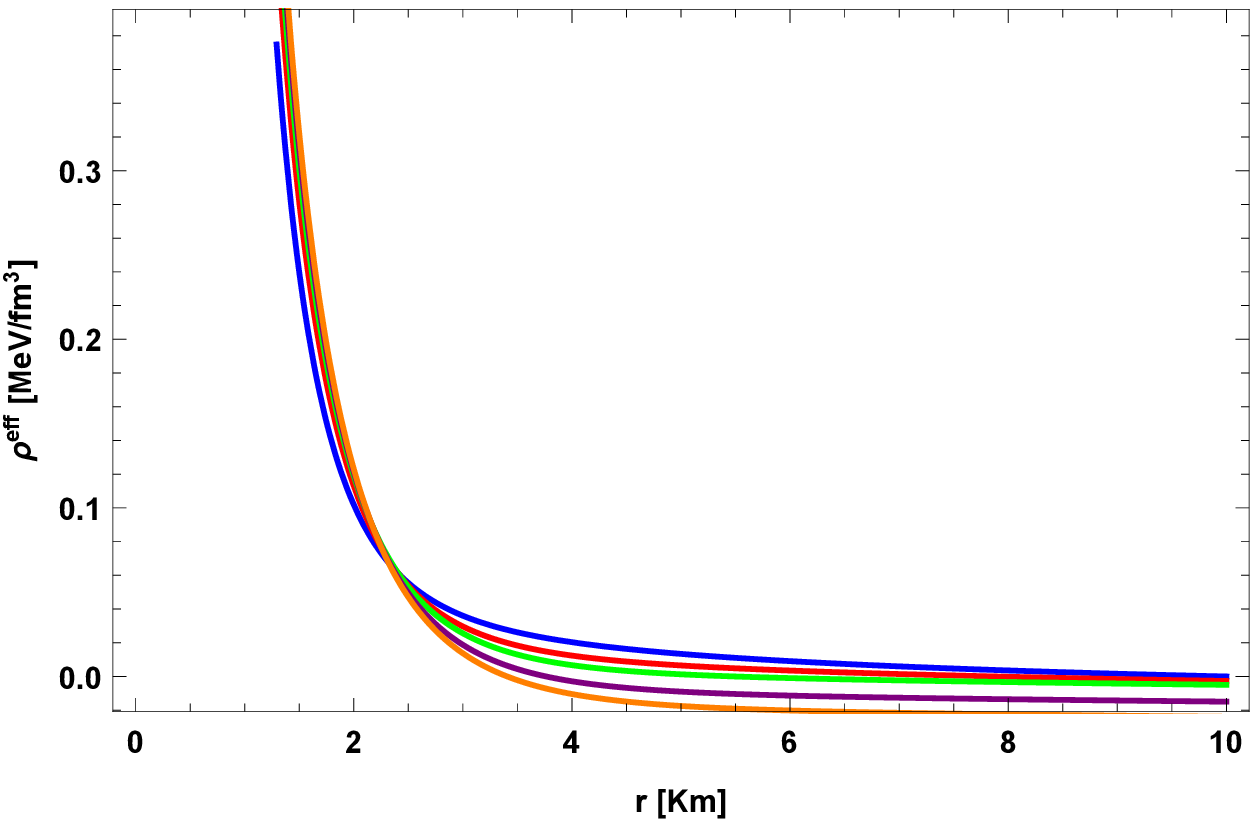,width=0.3\linewidth} &
\epsfig{file=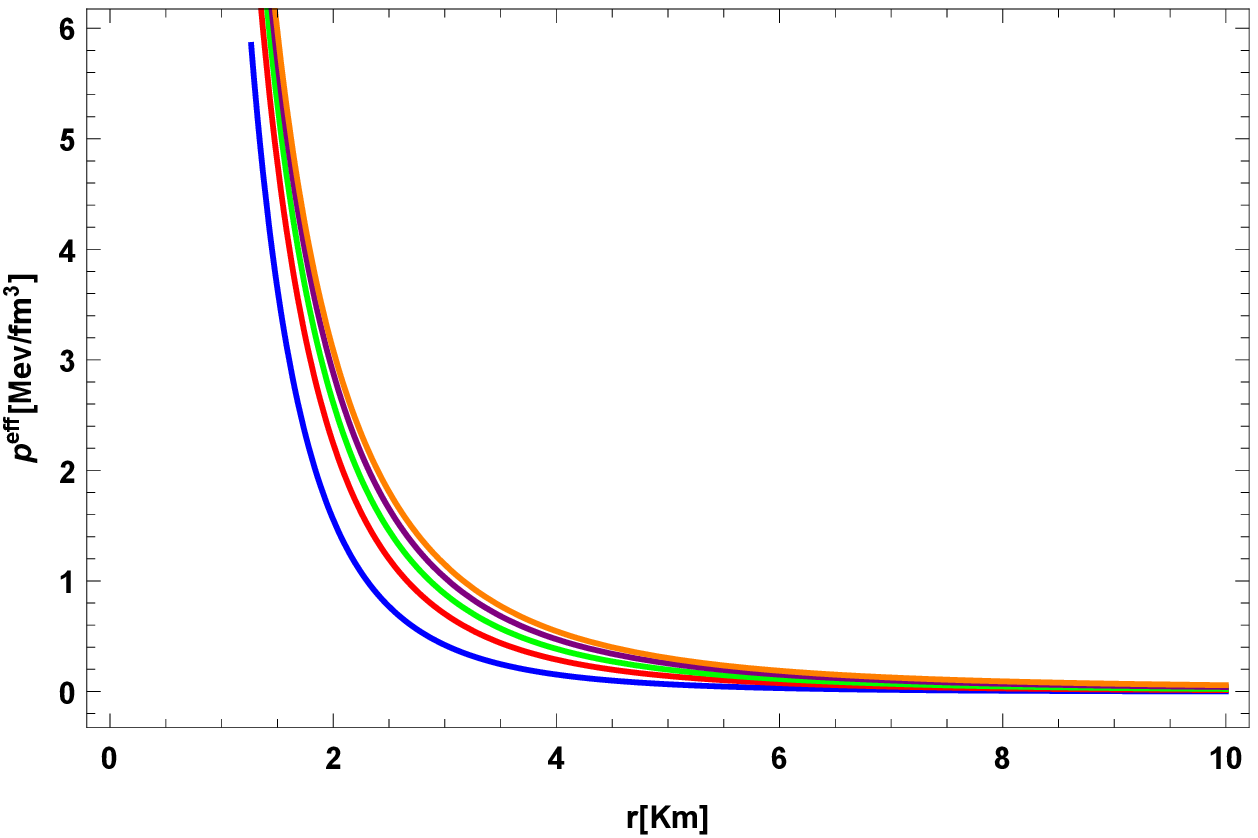,width=0.3\linewidth} &
\end{tabular}
\caption{Behavior of energy density $\rho^{eff}$ and pressure $p^{eff}$ for model-I.}\center
\label{Fig:a1}
\end{figure}

\begin{figure}\center
\begin{tabular}{cccc}
\epsfig{file=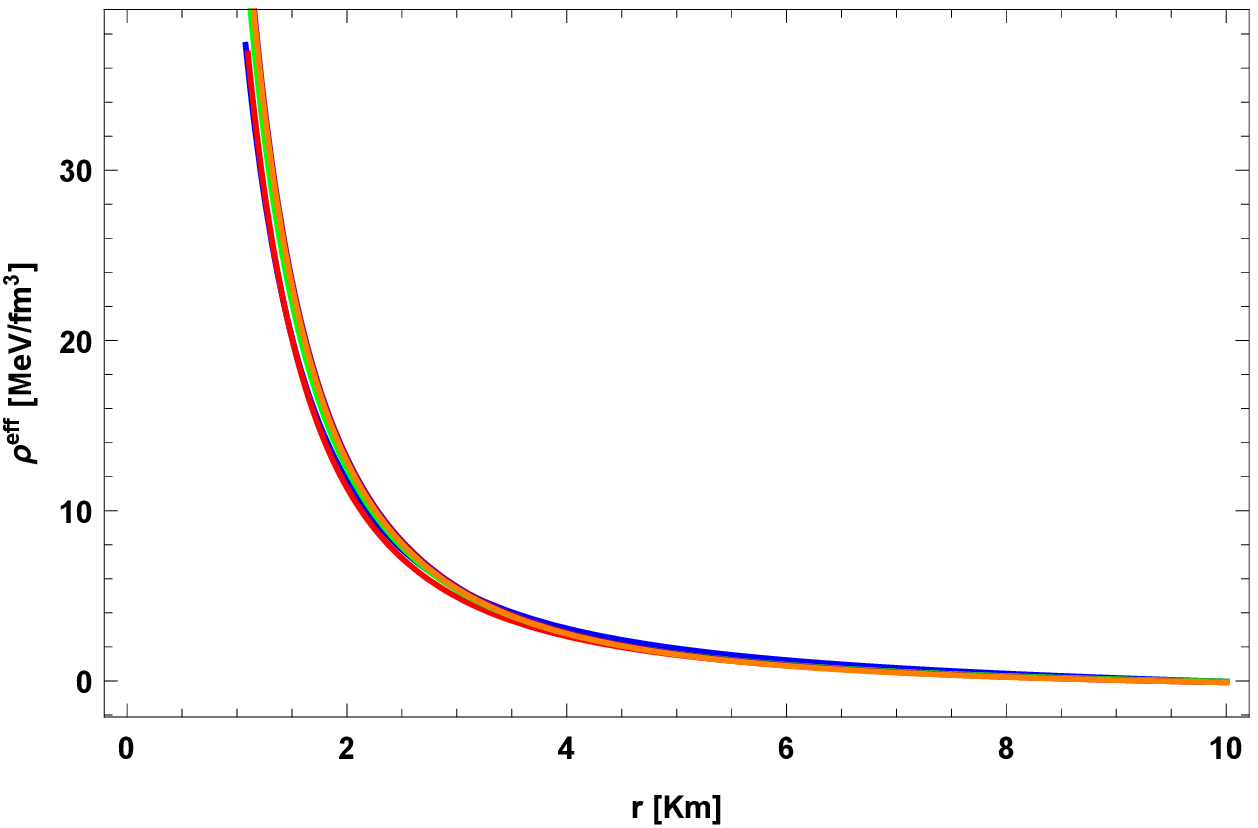,width=0.3\linewidth} &
\epsfig{file=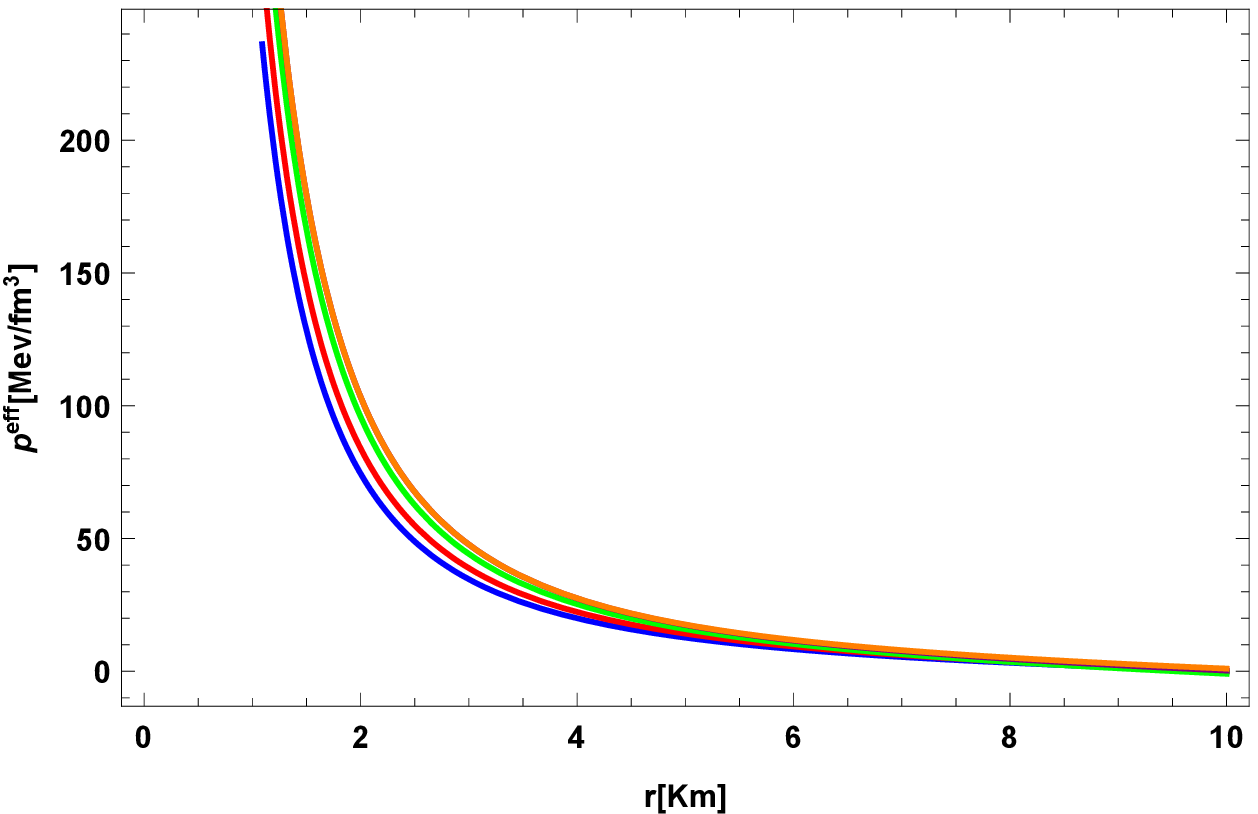,width=0.3\linewidth} &
\end{tabular}
\caption{Behavior of energy density $\rho^{eff}$ and pressure $p^{eff}$ for model-II.}\center
\label{Fig:a11}
\end{figure}

\begin{figure}\center
\begin{tabular}{cccc}
\epsfig{file=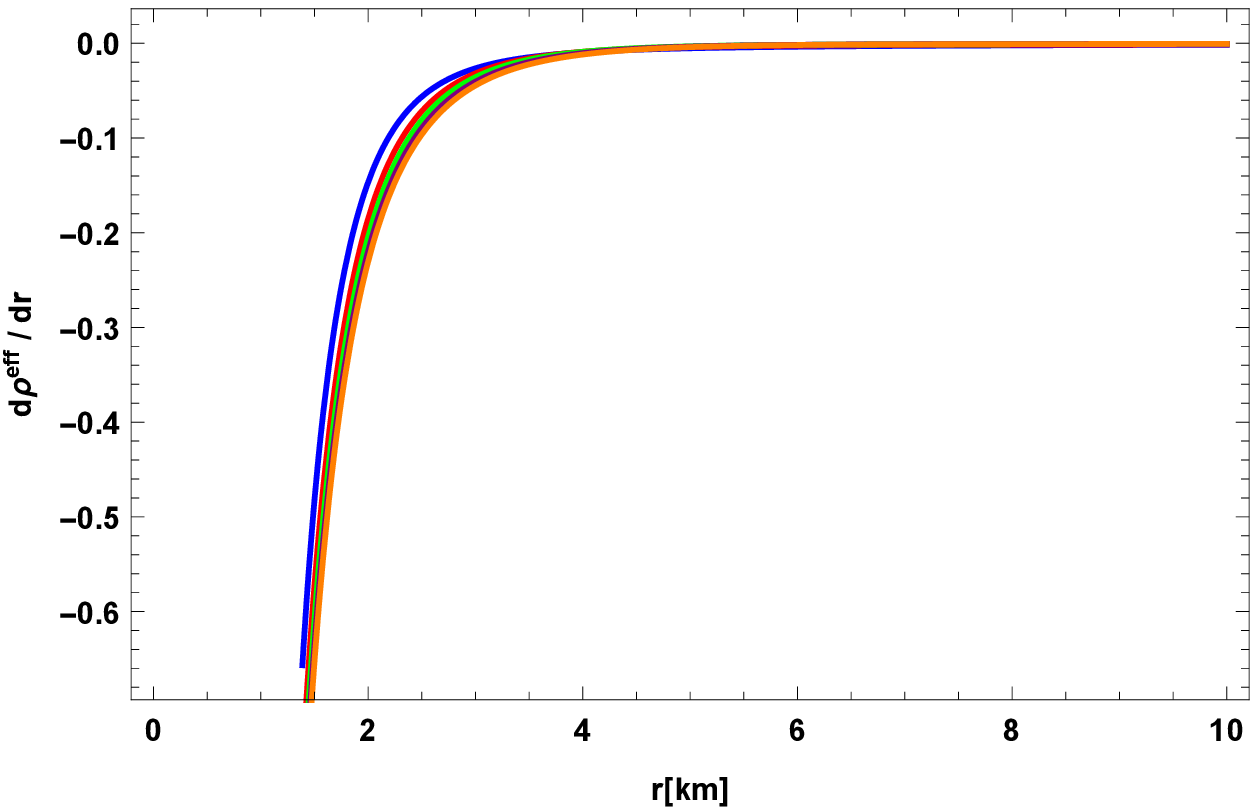,width=0.3\linewidth} &
\epsfig{file=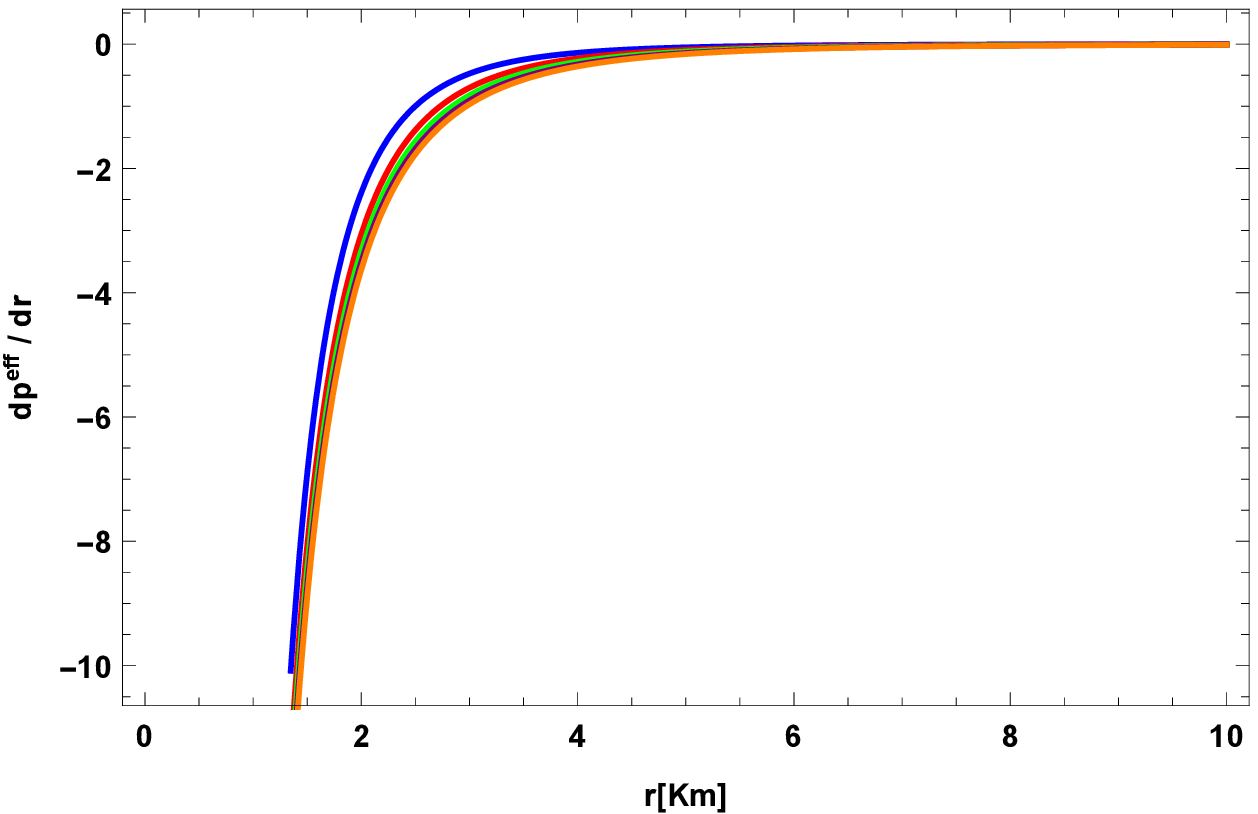,width=0.3\linewidth} &
\end{tabular}
\caption{The gradient behavior of energy density $\rho^{eff}$ and pressure $p^{eff}$ for model-I.}\center
\label{Fig:a2}
\end{figure}

\begin{figure}\center
\begin{tabular}{cccc}
\epsfig{file=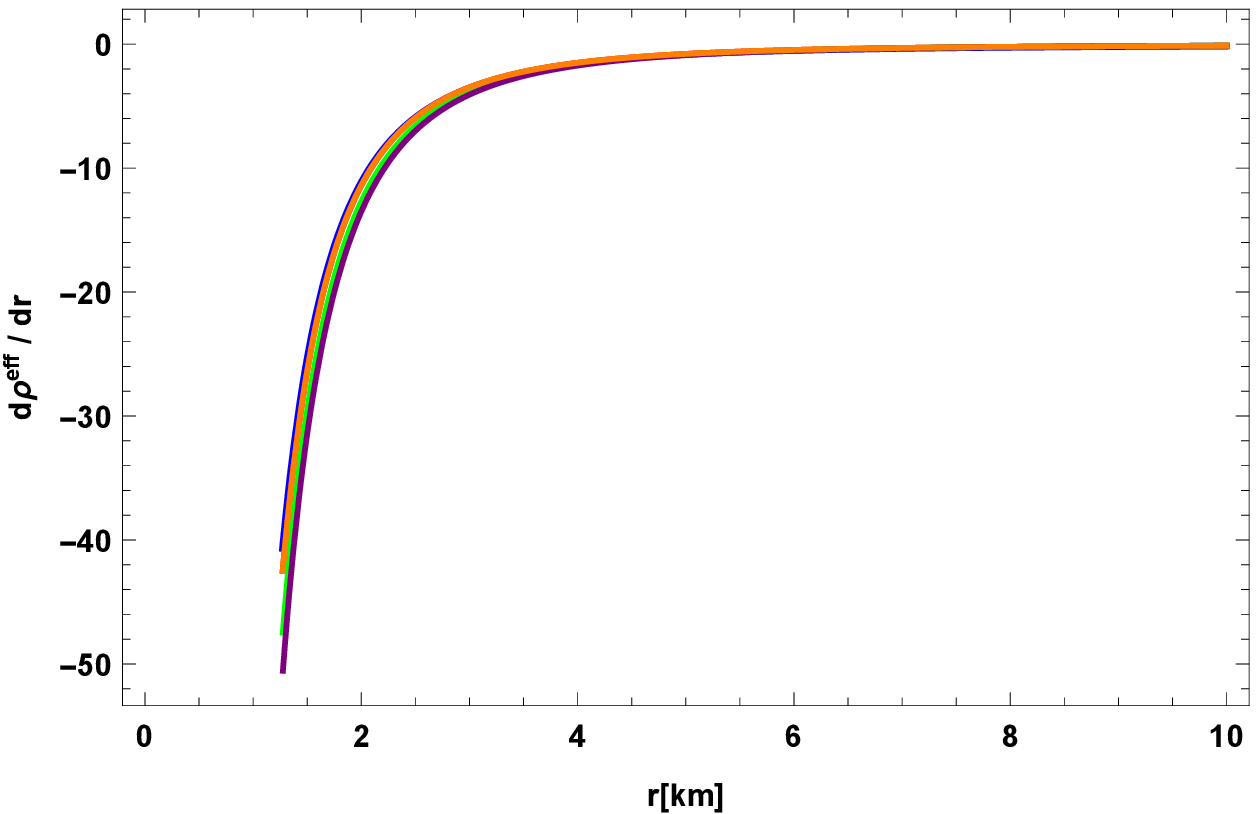,width=0.3\linewidth} &
\epsfig{file=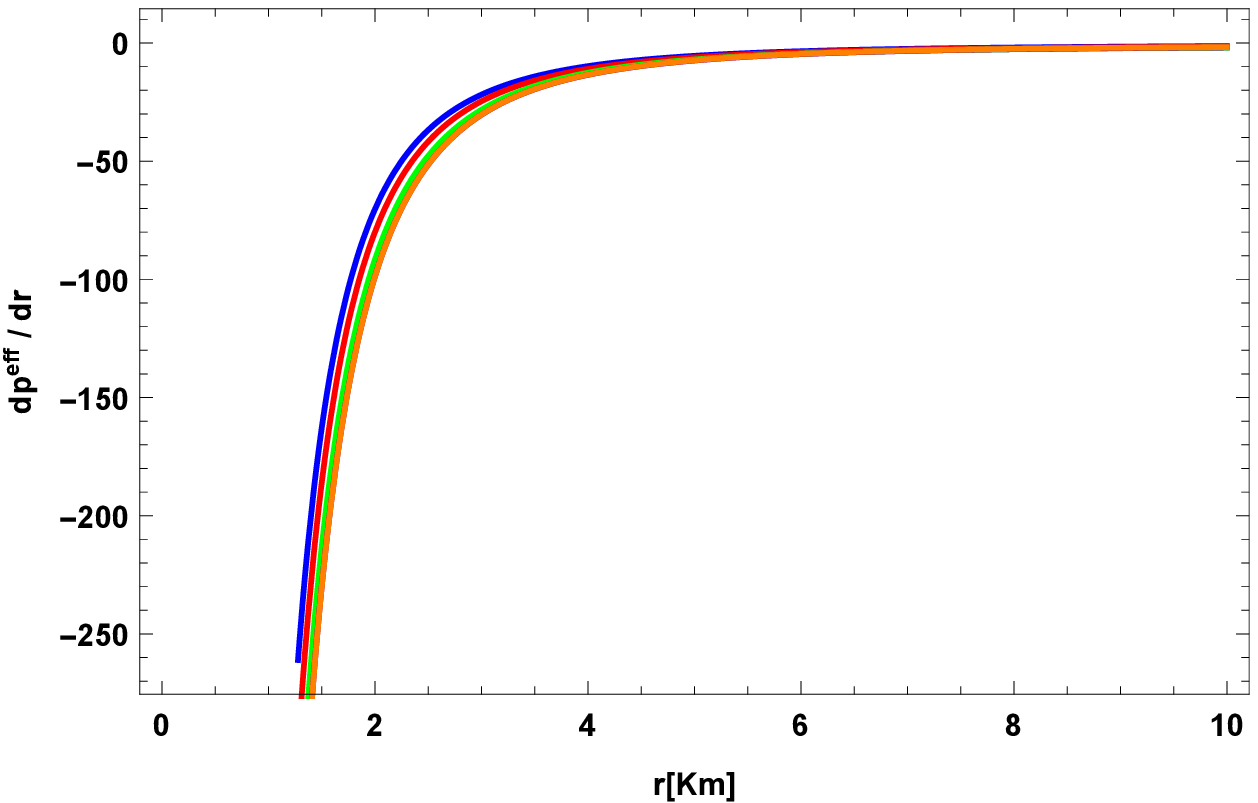,width=0.3\linewidth} &
\end{tabular}
\caption{The gradient behavior of energy density $\rho^{eff}$ and pressure $p^{eff}$ for model-II.}\center
\label{Fig:a22}
\end{figure}

Here, the energy density ($\rho^{eff}$) and pressure ($p^{eff}$) behave realistically for modified $f(R)$ gravity are show in Fig(\ref{Fig:a1}) for model-I and Fig(\ref{Fig:a11}) for model-II, which are decreasing and positive. The pressure is approaching to zero near the boundary of the star $r=R_b$. The concave up graphs of $\rho^{eff}$ and $p^{eff}$  are because of conformal motion and due to occurrence electric charge. The derivatives of energy density $\frac{d\rho}{dr}$ and  pressure $\frac{dp^{eff}}{dr}$ are shown in Fig(\ref{Fig:a2}) and Fig(\ref{Fig:a22}) for both models. The negative behavior shows that they are physically acceptable.

\subsection{Energy Conditions}
\begin{figure}\center
\begin{tabular}{cccc}
\epsfig{file=r0m1.eps,width=0.3\linewidth} &
\epsfig{file=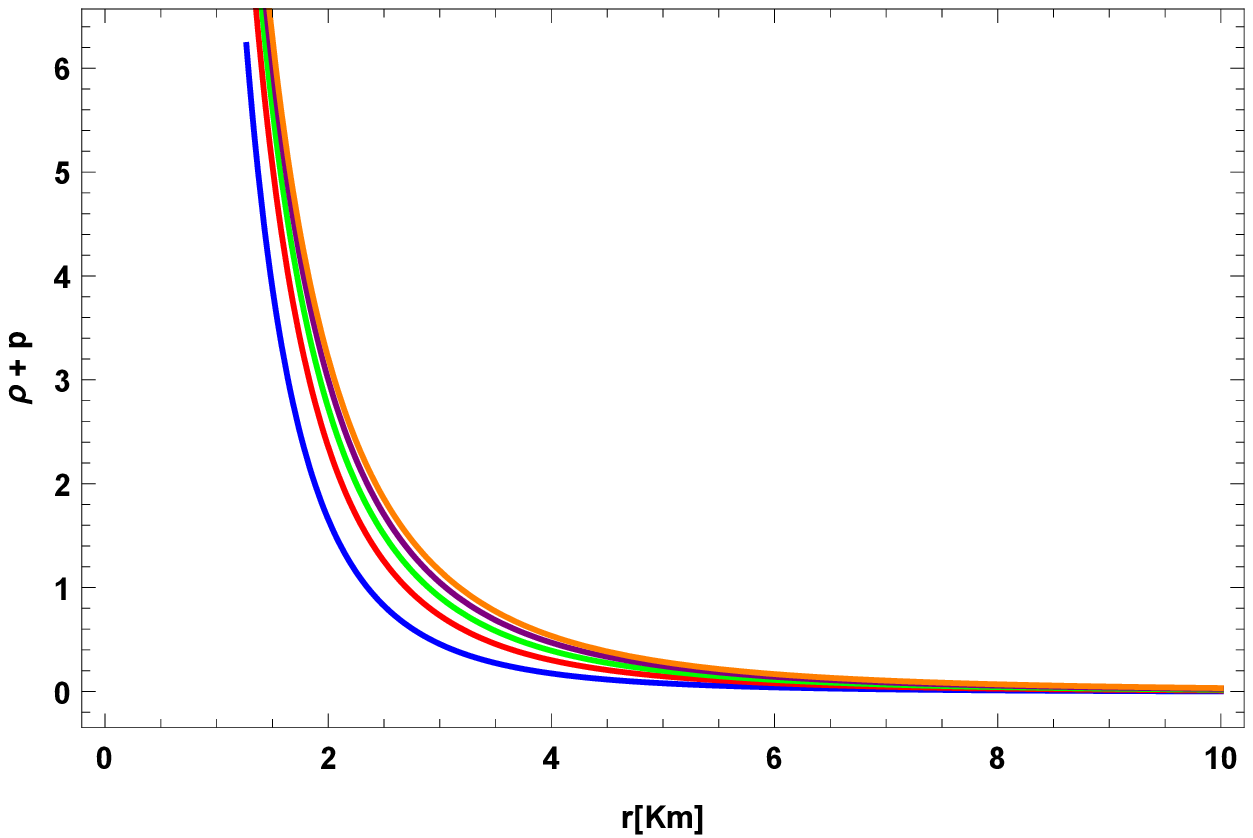,width=0.3\linewidth} &
\epsfig{file=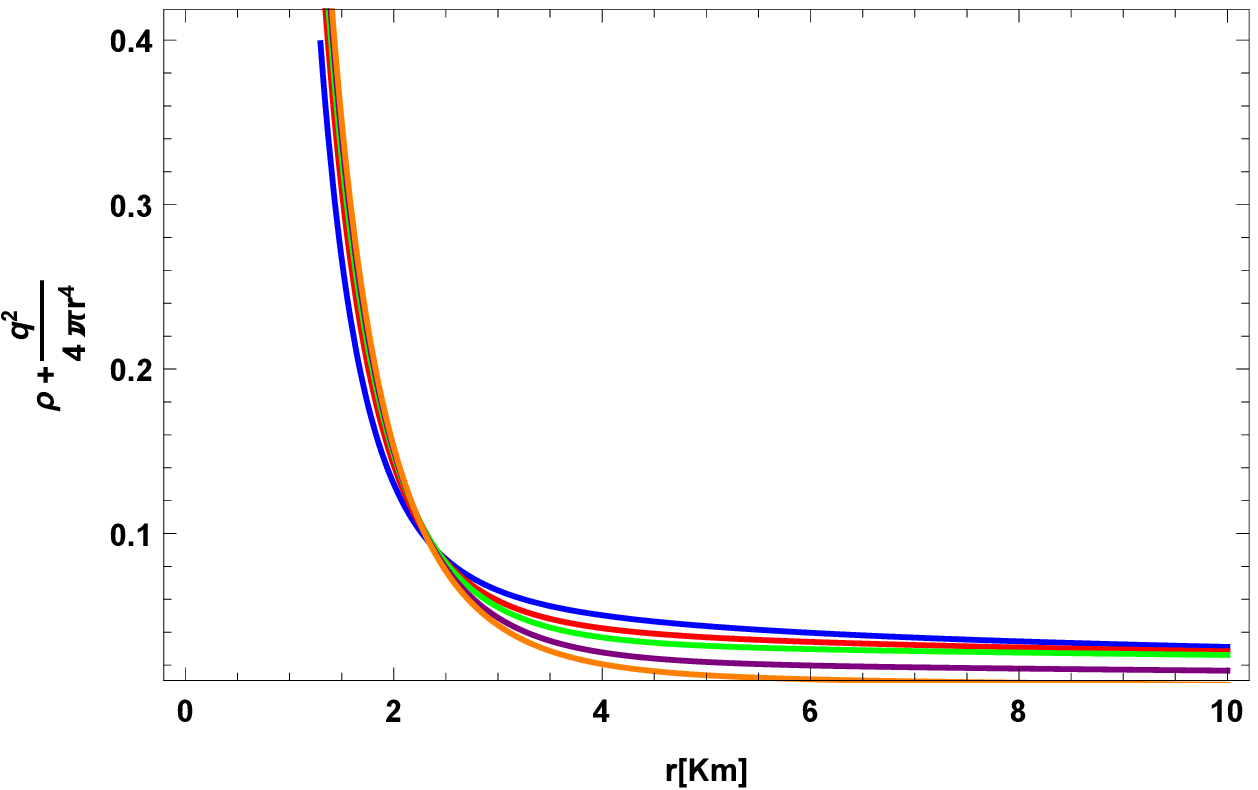,width=0.3\linewidth} &
\end{tabular}
\begin{tabular}{cccc}
\epsfig{file=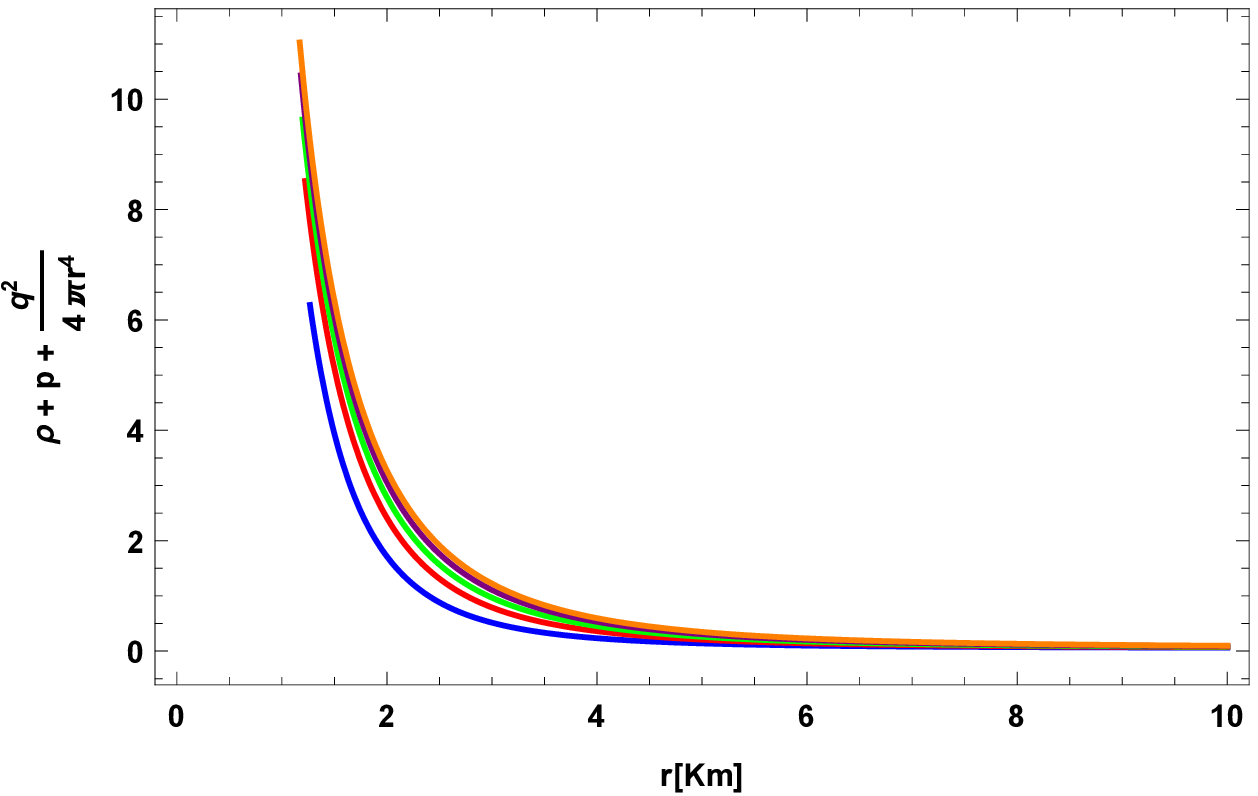,width=0.3\linewidth} &
\epsfig{file=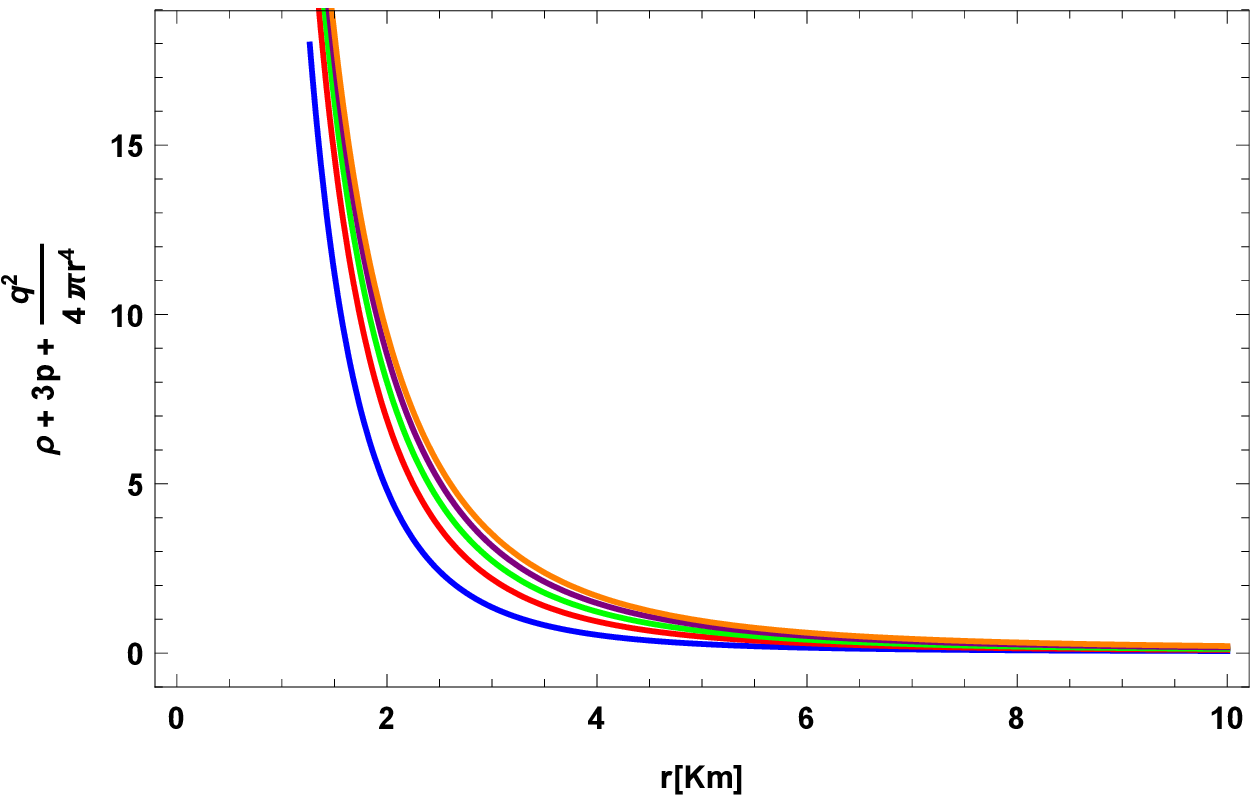,width=0.3\linewidth} &
\end{tabular}
\caption{Behavior of energy conditions for model-I.}\center
\label{Fig:a4}
\end{figure}

\begin{figure}\center
\begin{tabular}{cccc}
\epsfig{file=rom2.eps,width=0.3\linewidth} &
\epsfig{file=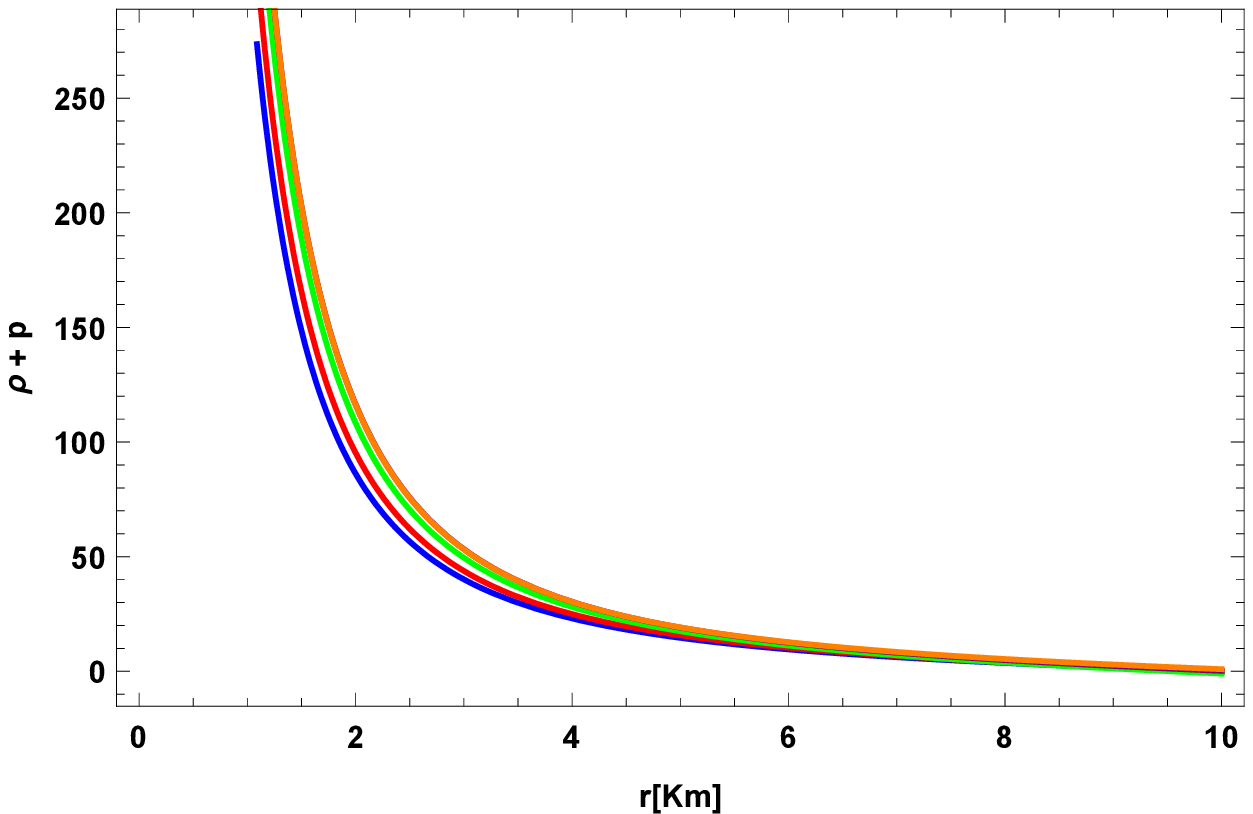,width=0.3\linewidth} &
\epsfig{file=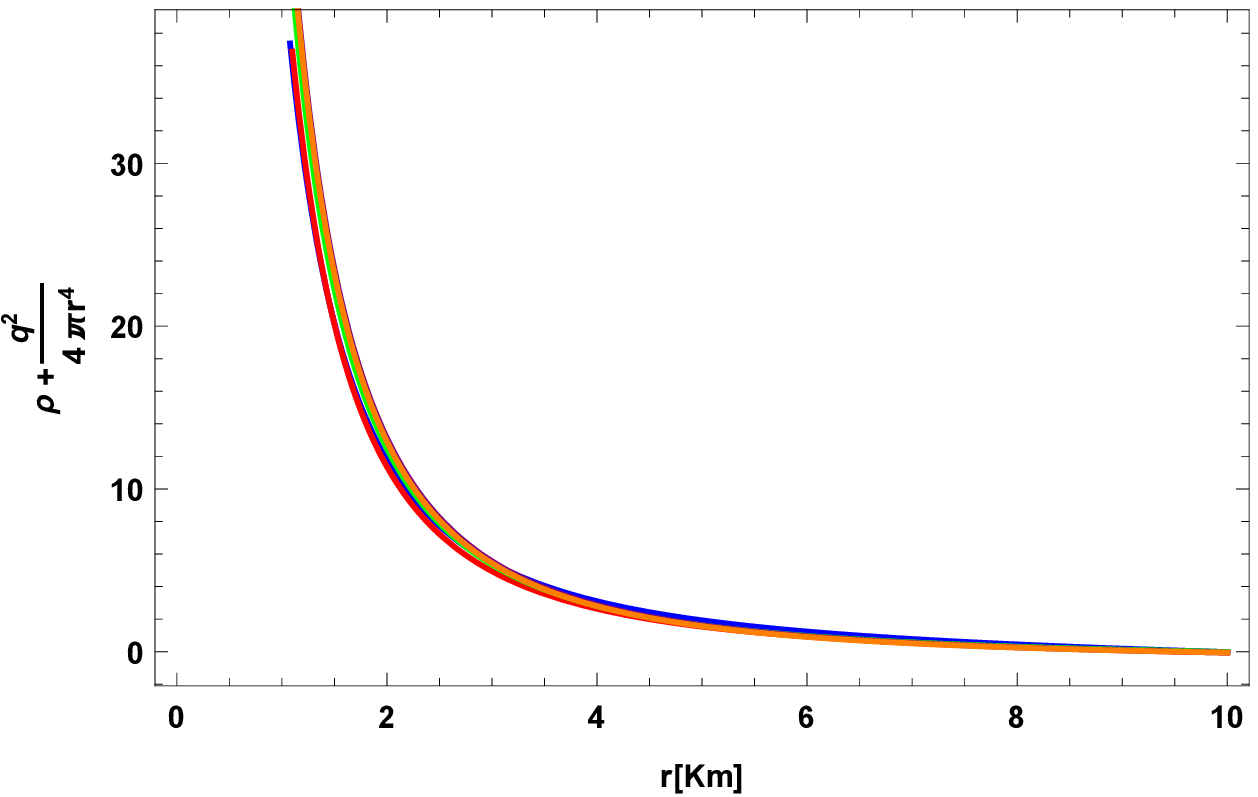,width=0.3\linewidth} &
\end{tabular}
\begin{tabular}{cccc}
\epsfig{file=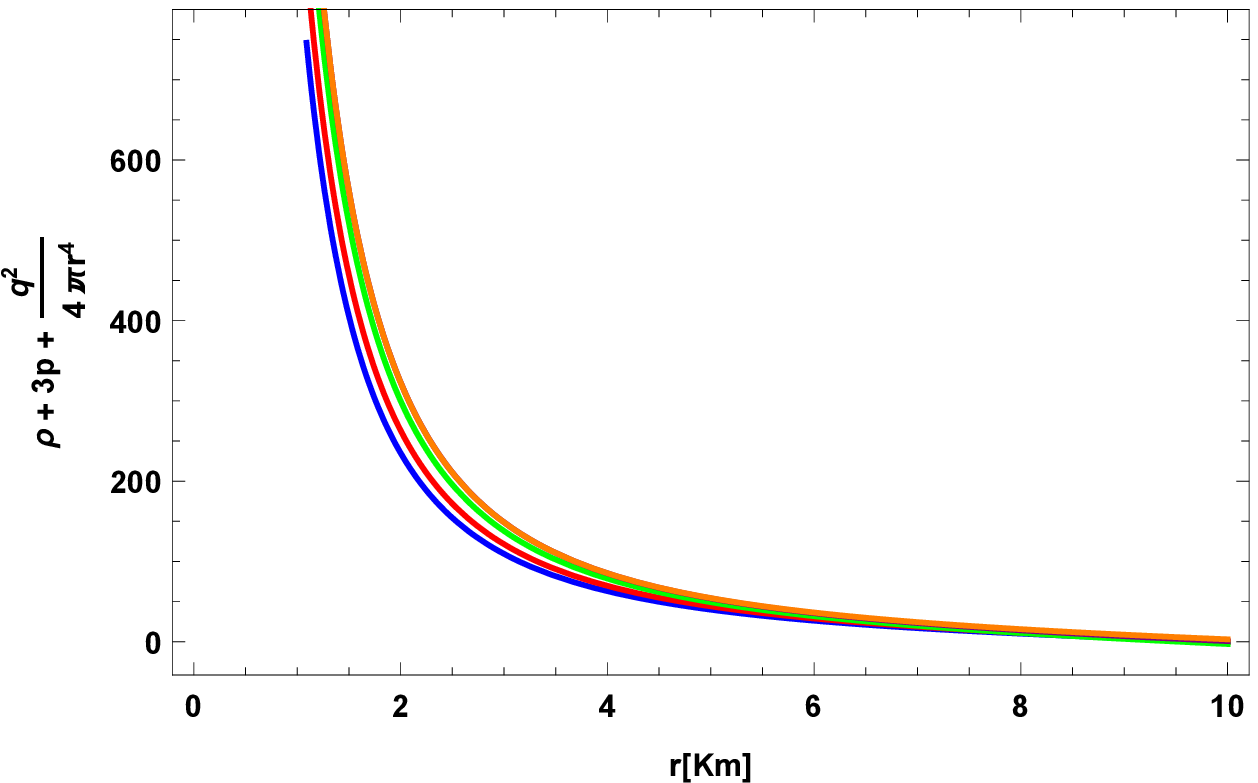,width=0.3\linewidth} &
\epsfig{file=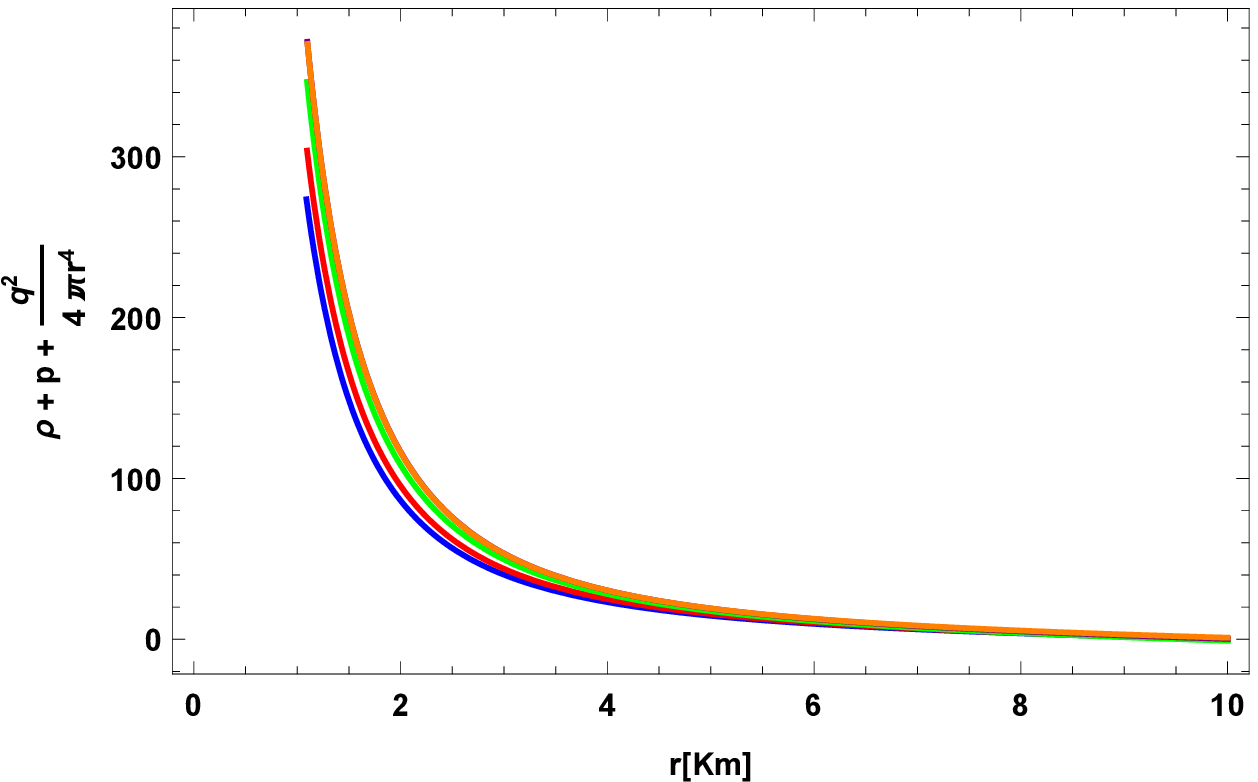,width=0.3\linewidth} &
\end{tabular}
\caption{Behavior of energy conditions for model-II.}\center
\label{Fig:a44}
\end{figure}

In cosmology the energy conditions have foremost importance to show the viability of the presented model. Many researchers have discussed the great features of energy conditions. It is  to satisfy the null energy conditions (NEC),weak energy condition (WEC), strong energy condition (SEC) and dominant energy condition (DEC). These energy conditions are given  by
\begin{itemize}

\item{Null energy conditions}
\begin{equation}
\rho^{eff}+ p^{eff} \geq 0,~~   \rho^{eff} +\frac{q^2}{4 \pi r^4} \geq 0,~~~~~~~~~~~~~~~~
\end{equation}
\item{Weak energy condition}
\begin{equation}
\rho+ \frac{q^2}{8 \pi r^4}\geq0,~~ \rho+ p\geq 0,~~  \rho + p +\frac{q^2}{4 \pi r^4} \geq 0,~~~~~~~~~~~~
\end{equation}
\item{Strong energy conditions}
\begin{equation}
\rho^{eff}+ p^{eff}\geq0,~~  \rho^{eff} + 3p^{eff} +\frac{q^2}{4 \pi r^4} \geq 0,~~~~~~~
\end{equation}
\end{itemize}
Above mentioned all energy conditions are satisfied for the selected $f(R)$ model as shown in Fig(\ref{Fig:a4}) and Fig(\ref{Fig:a44}).

\subsection{Equilibrium Conditions}
\begin{figure}\center
\begin{tabular}{cccc}
\epsfig{file=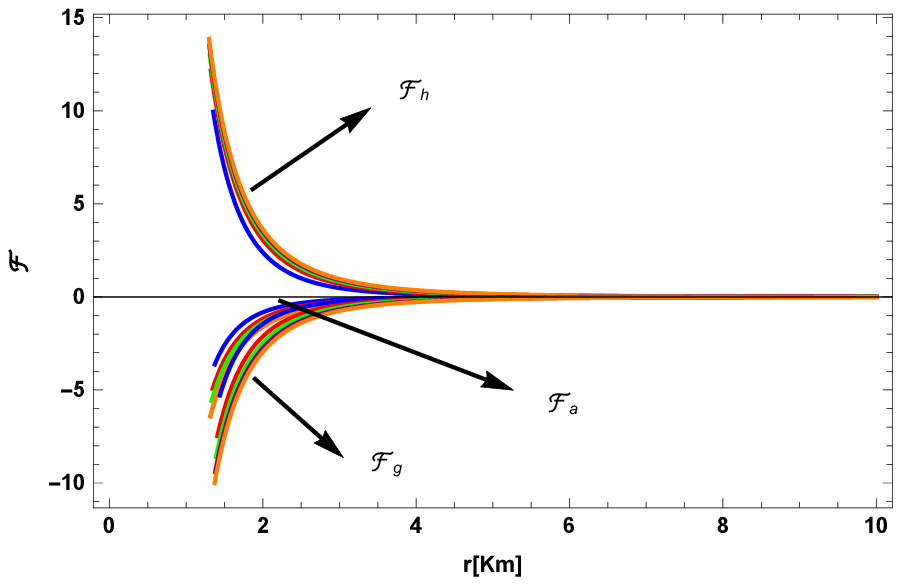,width=0.3\linewidth} &
\epsfig{file=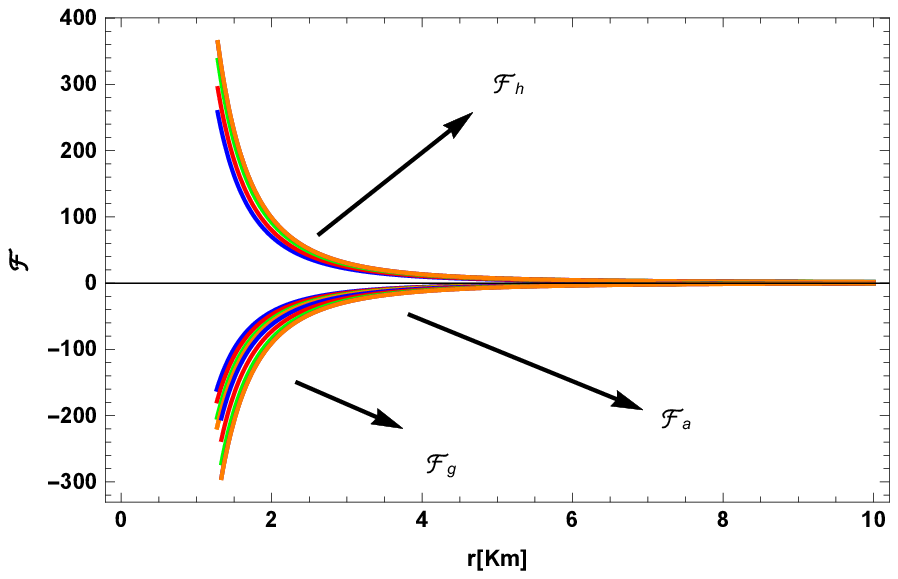,width=0.3\linewidth} &
\end{tabular}
\caption{Evolution of $\mathcal{F}_{h}$,$ \mathcal{F}_{g}$ and $\mathcal{F}_{e}$ for both modeles}\center
\label{Fig:a5}
\end{figure}

Here in this section, we discussed the equilibrium conditions in the scenario of Bardeen geometry with conformal motions for the existing charged stellar structure. The Tolman Oppenheimer Volkof $(TOV)$ equation is very useful to examine the equilibrium conditions for stellar structure. Fig(\ref{Fig:a5}) shows that all equilibrium conditions are satisfied for the current $f(R)$ models. The TOV equations for the charged isotropic fluid is defined as
\begin{equation}\label{29}
-\dfrac{\chi^{'}}{2}(\rho^{eff}+p^{eff})+\sigma(r)E(r)e^{\frac{\lambda(r)}{2}}-\dfrac{dp^{eff}}{dr}=0
\end{equation}
Above equation can be written as
\begin{equation}\label{30}
\mathcal{F}_{g}+\mathcal{F}_{e}+\mathcal{F}_{h}=0
\end{equation}

\begin{itemize}
\item{Gravitational Force}
\begin{equation}
\mathcal{F}_{g}=-\dfrac{\chi^{'}}{2}(\rho^{eff}+p^{eff})
\end{equation}
\item{Electric Force}
\begin{equation}
\mathcal{F}_{e}=\sigma(r)E(r)e^{\frac{\lambda(r)}{2}}
\end{equation}
\item{Hydrostatic Force}
\begin{equation}
\mathcal{F}_{h}=-\dfrac{dp^{eff}}{dr}
\end{equation}
\end{itemize}

\subsection{Equation of State Parameter}
\begin{figure}\center
\begin{tabular}{cccc}
\epsfig{file=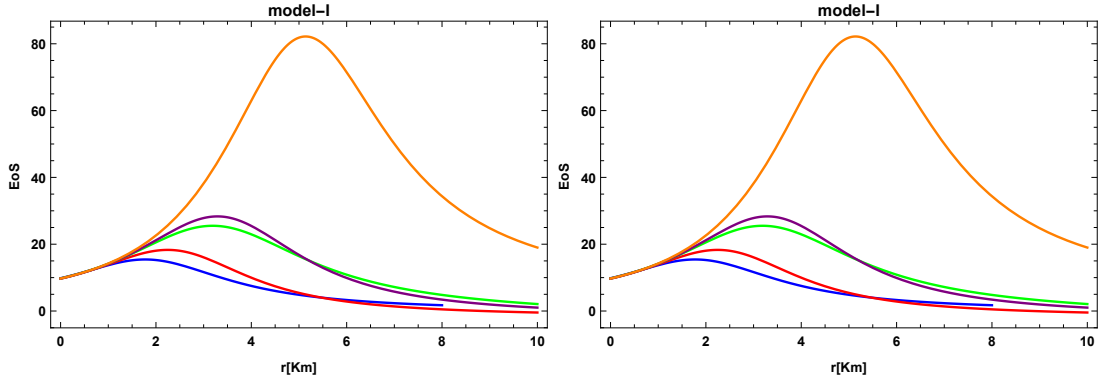,width=0.4\linewidth} &
\epsfig{file=Eosm1.eps,width=0.4\linewidth} &
\end{tabular}
\caption{Graph of equation of state  parameter $w$ for model I and II.}\center
\label{Fig:a6}
\end{figure}

In this section we presented the analysis of equation of state (EoS) parameters. The equation is given by the following expression,
\begin{equation}\label{311}
w=\dfrac{p^{eff}}{\rho^{eff}}.
\end{equation}
The graphical behavior of is given in Fig(\ref{Fig:a6}), the graph of model-I shows that for radius less than $8$, the effective pressure is increasing, this means these equations of state is often referred to as finite strain theory to compare their behavior from the predictions of elastic strain theory, in which the magnitude of volumetric strains are assumed to be infinitesimal as described in \cite{corm}. Where as near outer boundary the equation of state is within the range which shows that the star consists of the normal matter, stable and indicates the high compactness at the outer boundary.
%
%
%
%

\subsection{Adiabatic Index}
\begin{figure}\center
\begin{tabular}{cccc}
\epsfig{file=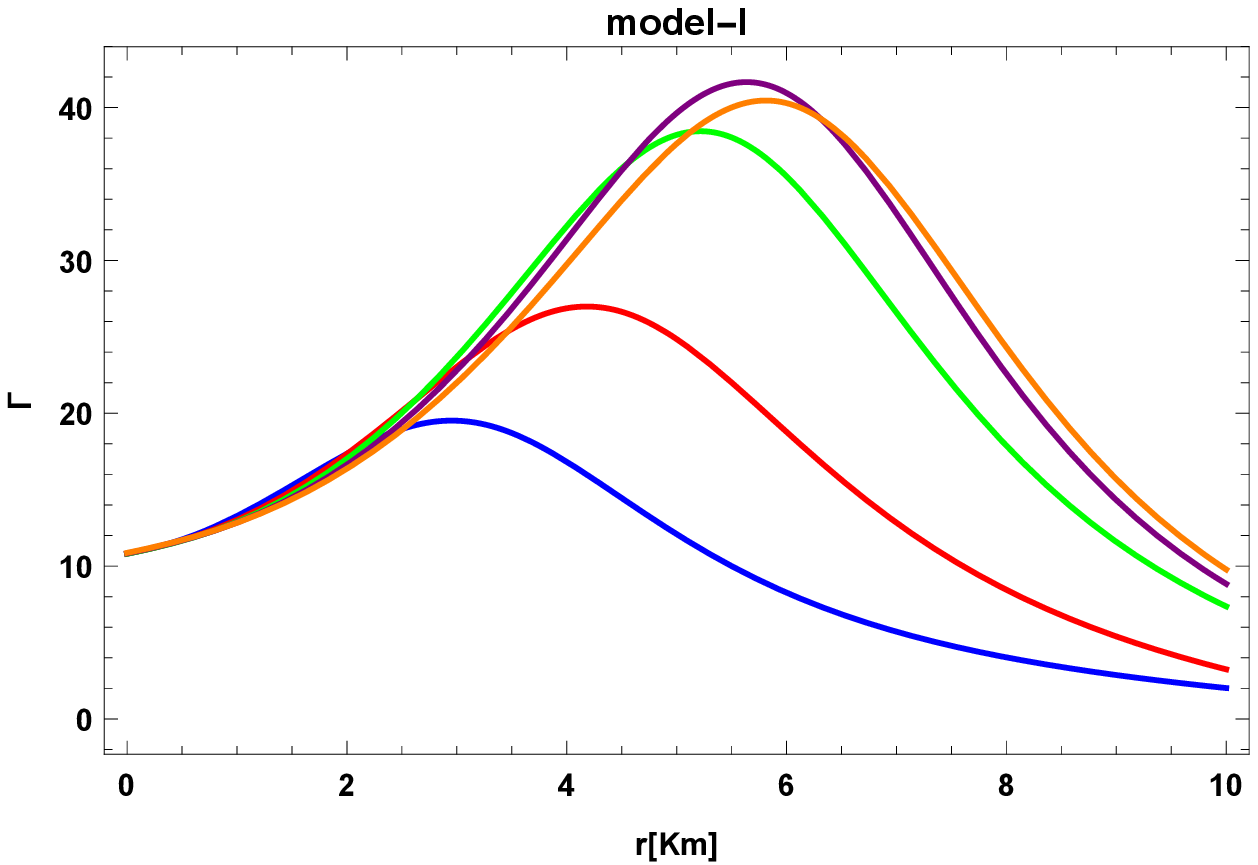,width=0.4\linewidth} &
\epsfig{file=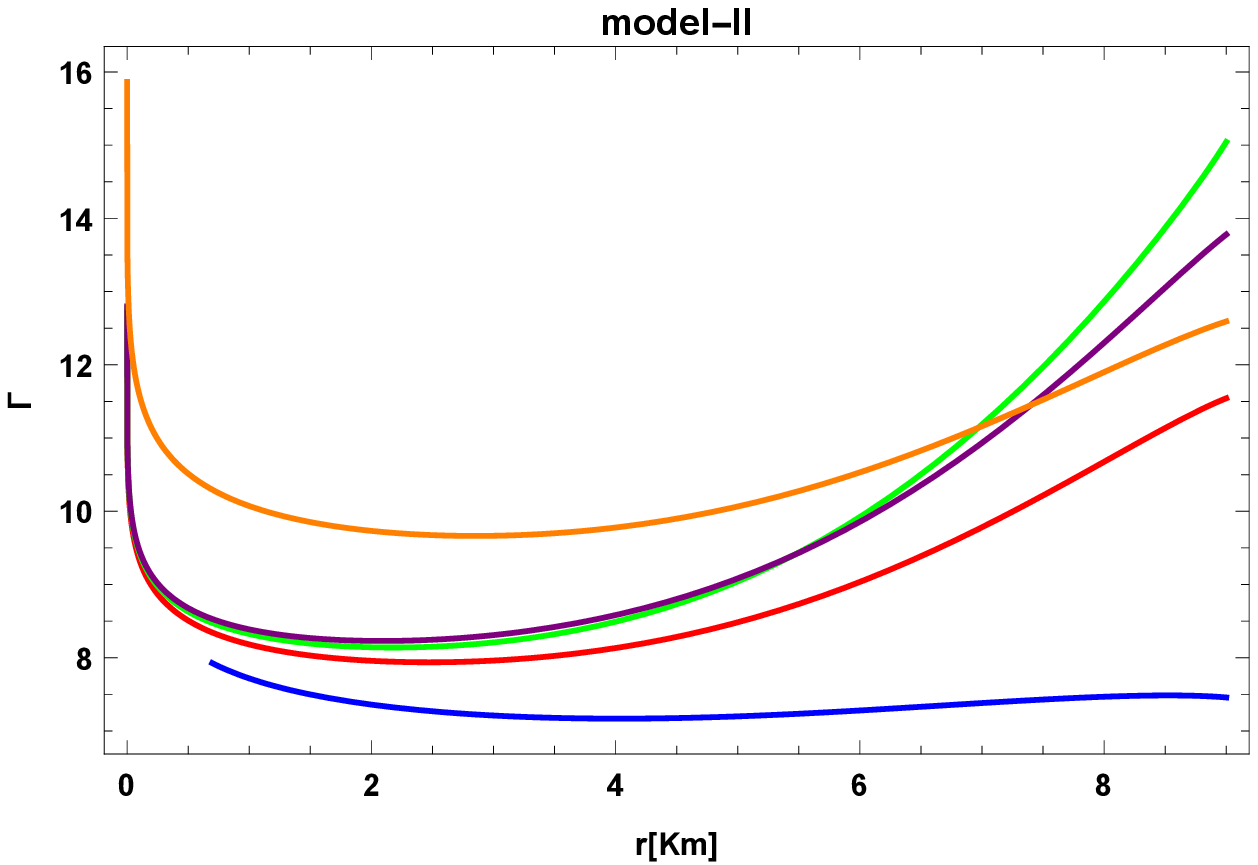,width=0.4\linewidth} &
\end{tabular}
\caption{Evolution of adiabatic index.}\center
\label{Fig:a8}
\end{figure}

The adiabatic index has a key factor that reflects the stiffness of equation of state. Chandrasekhar \cite{chnd} presented insatiability criteria depending upon adiabatic index, which is a good separator between a large gravitational field and a small abhorrent nuclear forces. The formulation of adiabatic index is given by
\begin{equation}
\Gamma=(1+\dfrac{\rho^{eff}}{p^{eff}}\dfrac{dp^{eff}}{d\rho^{eff}}).
\end{equation}
The adiabatic index must be greater than $4/3$, Fig(\ref{Fig:a8}) shows that the our models are stable.

\subsection{Mass-Radius Relation, Compactness Factor, Surface Red-Shift}

The mass-radius relation for a compact star must be within the limits i.e. $2M/R \leq 8/9$ \cite{buchd}
\begin{equation}\label{322}
M(r)=4\pi\int_{0}^{\pi}\rho^{eff}r^{2}dr=\dfrac{R}{2}(1-e^{\lambda}+\frac{q^2}{r^2})
\end{equation}

For charged compact stars the mass function have to fulfill Andreasson's limit \cite{and} given as $\sqrt{M}\leq\frac{\sqrt{R_b}}{3}+\sqrt{\frac{R_b}{9}+\frac{q^2}{3 R_b}}$. The details are given in table \ref{tab:3}.
\begin{table}[ht]
\caption{Estimated values of $M$, $R_b$ and  Andreasson's limit for model-I and model-II.}
\centering
\begin{tabular}{|p{2.3cm}|p{1.8cm}| p{2.2cm}| p{2.2cm}| p{2.3cm}|}
\hline\hline
$M(M_{\odot})$ & $R_b$ & $\sqrt{M}$ & Model-I & Model-II \\
\hline
$2.97979661$ & $9.12904$  & 1.665850658 & 1.92155  & 2.57812  \\
\hline
$2.97644068$ & $9.14507$  &1.65850249 & 1.91807 & 2.59772   \\
\hline
$2.96298998$ & $9.20902$  & 1.65841051 & 1.91719 & 2.64241 \\
\hline
$2.94349831$ & $9.30119$  & 1.65806095 & 1.91338 & 2.66736 \\
\hline
$2.93093559$ & $9.36021$  & 1.65770708 & 1.91384 & 2.63862 \\
\hline
\end{tabular}
\label{tab:3}
\end{table}

\begin{figure}\center
\begin{tabular}{cccc}
\epsfig{file=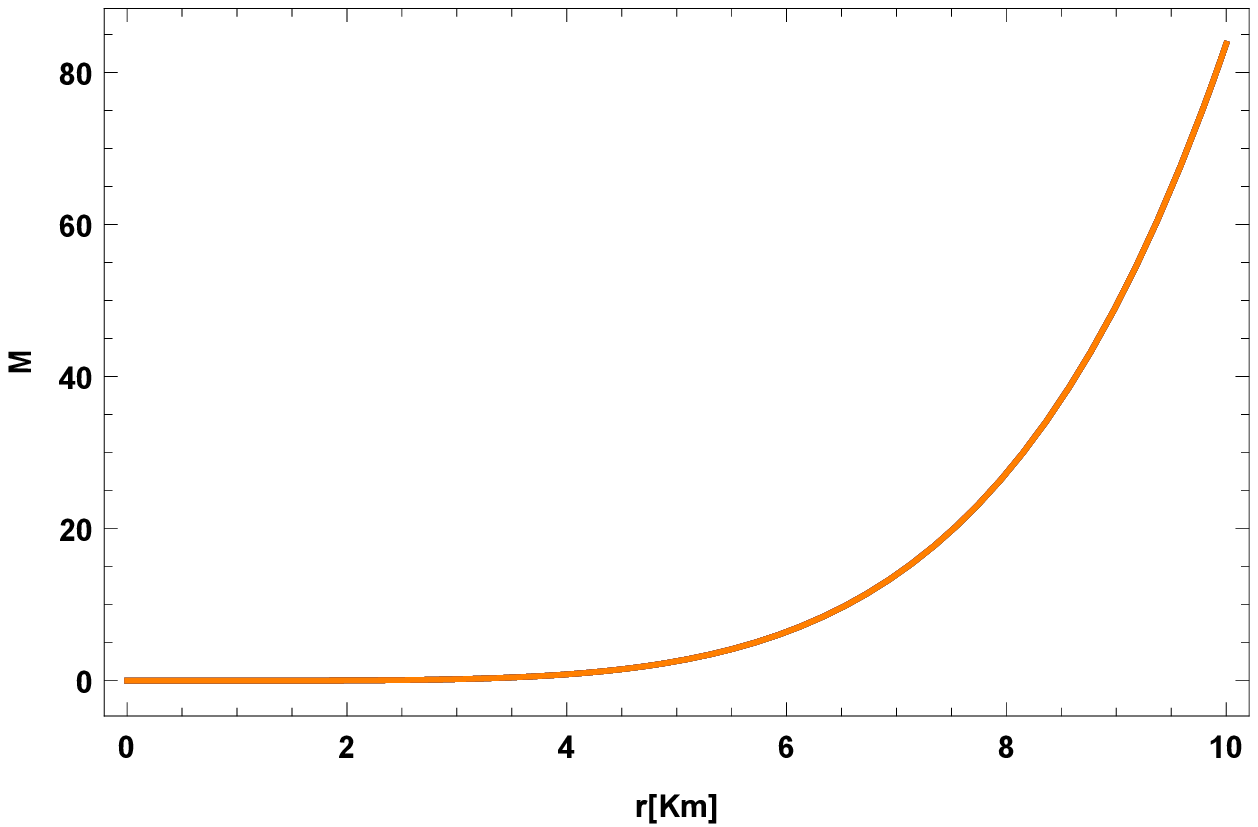,width=0.3\linewidth} &
\epsfig{file=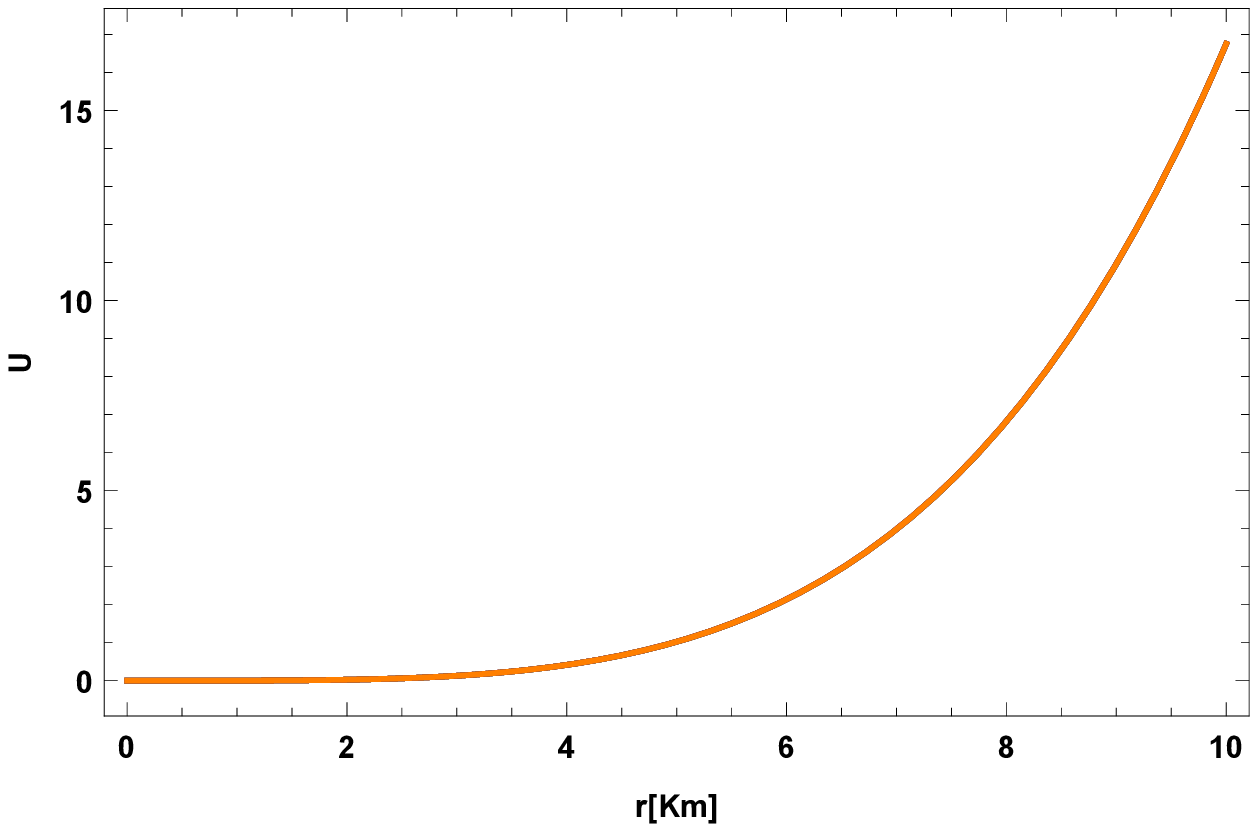,width=0.3\linewidth} &
\epsfig{file=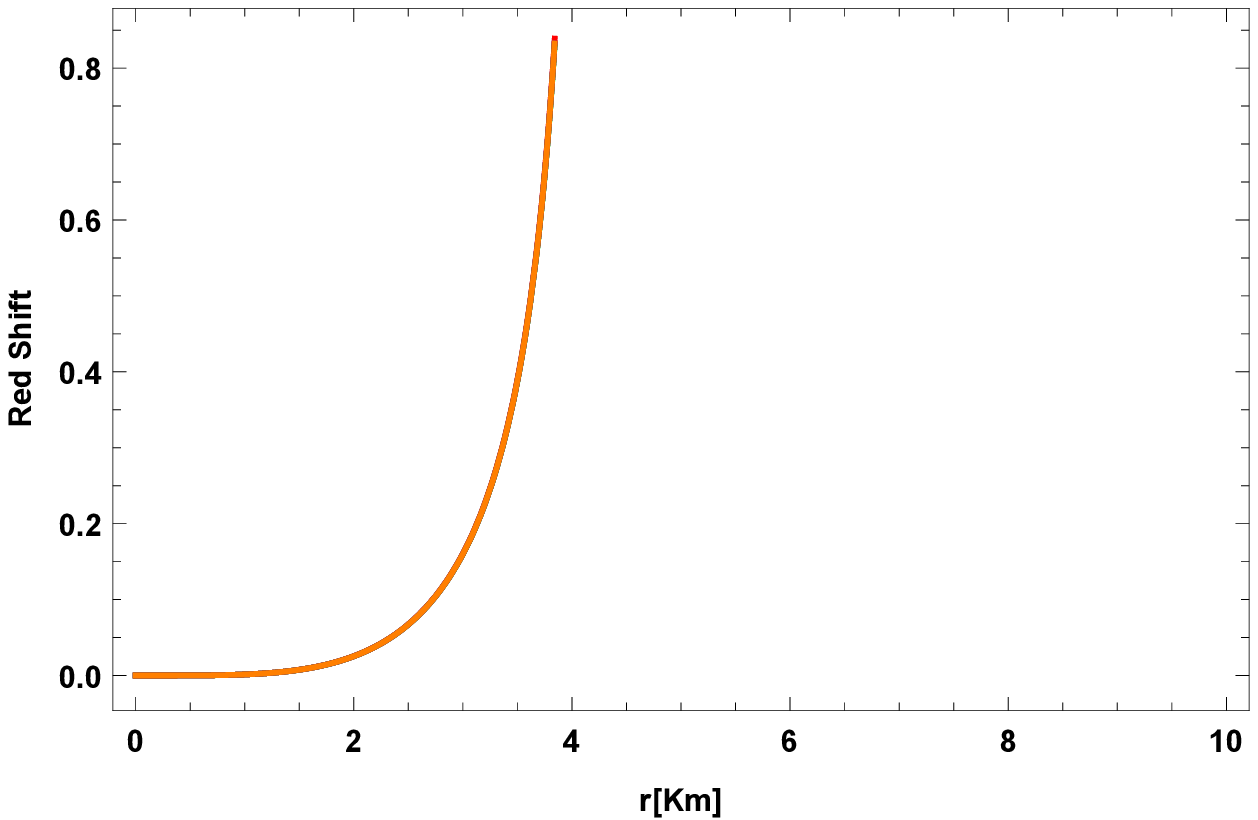,width=0.3\linewidth} &
\end{tabular}
\caption{Graphical behavior of mass function, compactness factor and red shifts for model-I.}\center
\label{Fig:a9}
\end{figure}

\begin{figure}\center
\begin{tabular}{cccc}
\epsfig{file=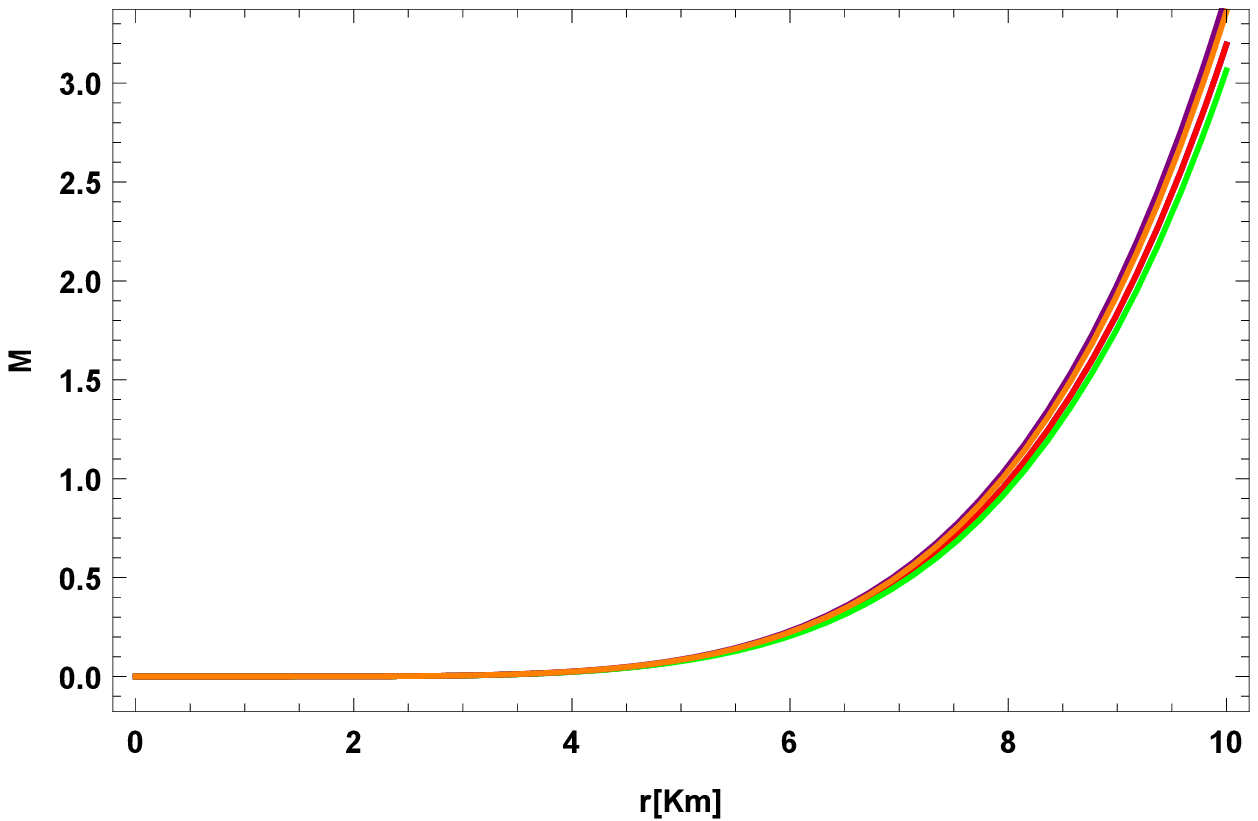,width=0.3\linewidth} &
\epsfig{file=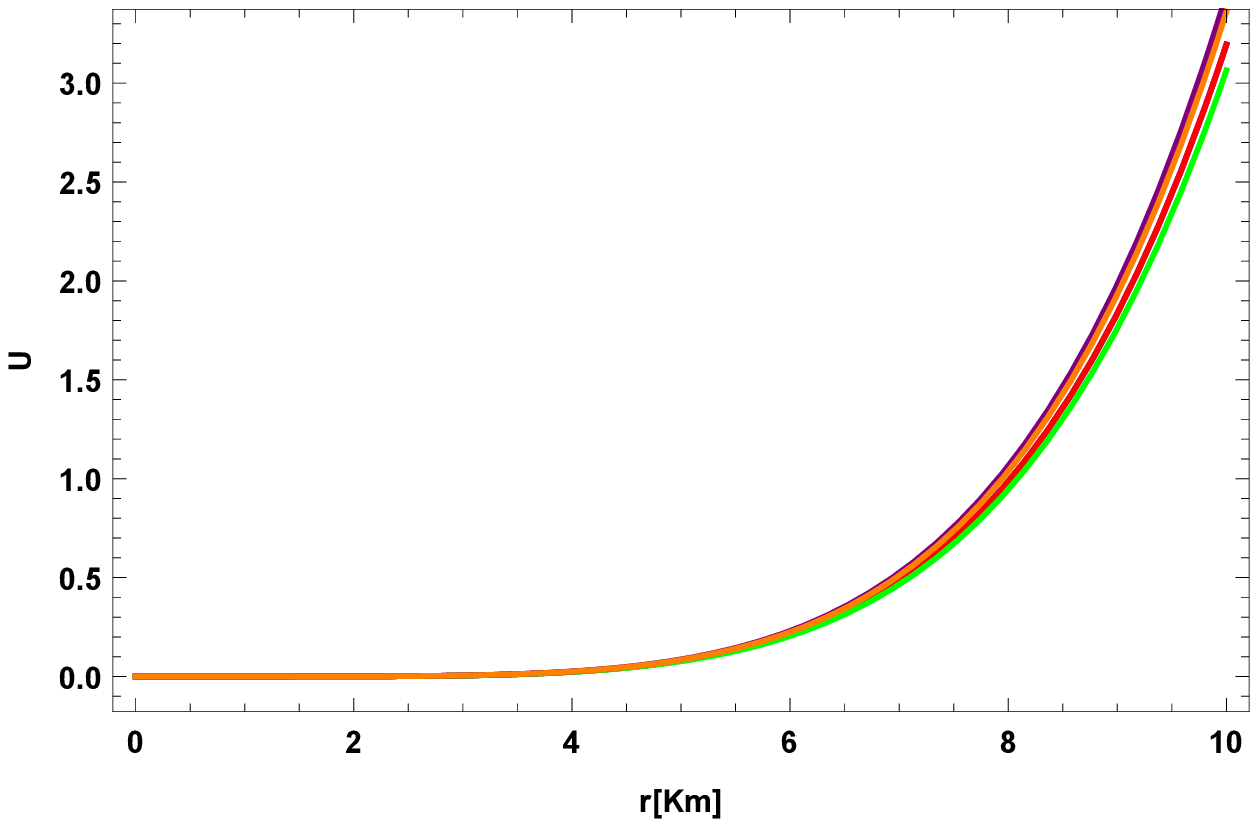,width=0.3\linewidth} &
\epsfig{file=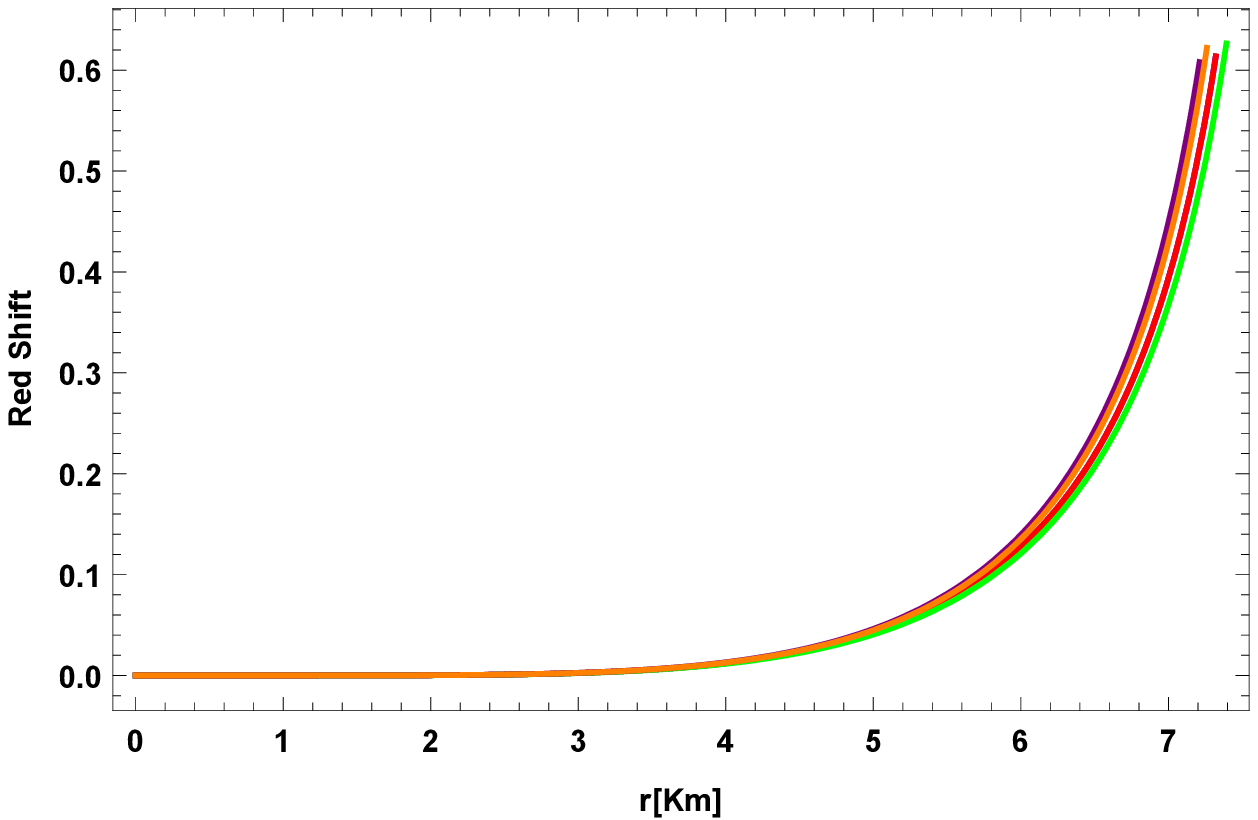,width=0.3\linewidth} &
\end{tabular}
\caption{Graphical behavior of mass function, compactness factor and re shifts for model-II.}\center
\label{Fig:a10}
\end{figure}

Here the compactness parameter \cite{iva,makh11} and surface red shift \cite{smh} are denoted by $U(r)$ and $Z_{rs}$ defined as
\begin{equation}\label{32}
U(r)=\frac{M}{r}=\frac{1-e^{-\lambda}}{2}+\frac{q^2}{2r^2},
\end{equation}
\begin{equation}\label{323}
  Z_{rs}=(1-U(r))^{-1/2}-1
\end{equation}

We have plotted the mass function, compactness parameter and surface red shift in Fig(\ref{Fig:a9}) and Fig(\ref{Fig:a10}) for model-I and II respectively. The graphical behavior of all these are monotonically increasing.

\section{Conclusion And Comparison}
In this paper, we select Bardeen black hole geometry with conformal motion and applied matching condition on two different models of modified $f(R)$ gravity  i.e. $f(R)=R+mR^2$ and $f(R)= R+ \alpha R^{2}+\beta R^{2}ln(\beta R)$ with isotropic energy momentum tensor in presence of electric charge. The analysis have been done for different values of $m$ which are  $m=0.10$, $m=0.2$, $m=0.3$, $m=0.4$, and $m=0.5$ for model I. For model II we fixed the $g$ and $\alpha$ and varied the value of $\beta$ to get viable results. The analysis of our work is enumerated below
\begin{itemize}
\item{The solutions found with conformal symmetries for the metric potentials are finite, bounded and singularity free across the star and also at the boundary from $r=0$ to $r=R_{b}$. The analysis of energy density, pressure are shown in Fig(\ref{Fig:a1}). This study is done with $R_{b}=9.12904$, $R_{b}=9.14507$, $R_{b}=9.20909$, $R_{b}=9.30119$ and $R_{b}=9,36021$. The derivatives $d\rho/dr$ and $dp/dr$  are negative which illustrate that they are acceptable for the present study.}

\item{We have presented the energy conditions in this paper which are WEC, NEC and SEC. All the energy conditions are satisfied for both models under conformal motion. The graphical illustration of all energy conditions are shown in Fig(\ref{Fig:a4}) and Fig(\ref{Fig:a44}).}

\item{For the existing charged stellar structure, we presented the equilibrium conditions by using TOV equations in the scenario of Bardeen geometry along with conformal motions are shown in Fig(\ref{Fig:a5}), the positive graphs shows that all equilibrium conditions are satisfied for the presented $f(R)$ models. The balancing development of the graphs of forces shows that our models are viable and stable in presence of electric charge through conformal motion with Bardeen geometry.}
\item{The adiabatic index is the main element which shows the stiffness of equation of state. Here in current scenario, the value of Adiabatic index is as per the requirement i.e. $\Gamma$ ie greater $4/3$ as given in \cite{chnd}.}

\item{It is observed that the model-I satisfied the equation of state parameters near the outer boundary which implies that the effective pressure is increasing and near outer boundary the equation of state is within the range which shows that the star consists of the normal matter, stable and indicates the high compactness at the outer boundary. The graph of model-II satisfied the limits i.e. these parameters must lie between 0 and 1.}

\item{The mass function satisfied the Andreasson's limit \cite{and} amd the details are given in table \ref{tab:3}. The compactness parameter and surface red-shift have monotonically increasing behaviour which shows the stability of our results. }

\end{itemize}

In the end, we have effectively conducted a stable and viable stellar structure under the presence of charge by using two different models of modified $f(R)$ gravity i.e. $f(R)=R+mR^{2}$ and $f(R)= R+ \alpha R^{2}+\beta R^{2}ln(\beta R)$ with Bardeen black hole geometry along with conformal motion. Conclusively, the derived calculated solutions demonstrate authentic evidences for better, realistic and viable stellar structure in presence of charge with conformal motion with Bardeen geometry.


\section*{References}

\begin{thebibliography}{70}

\bibitem{Bar} J. M. Bardeen, Non-singular general-relativistic gravitational collapse, Proceedings of GR-5, Tiflis, Georgia, U.S.S.R. page \textbf{174}(1968).

\bibitem{Fer} S. Fernando and J. Correa, Quasinormal modes of Bardeen black hole: scalar perturbations, \textit{Phys. Rev. D.} \textbf{86} (2012) 64039.

\bibitem{shj} M. Sharif and W. Javed, Quantum corrections for a Bardeen regular black hole, \textit{J. Korean Phys. Soc.} \textbf{57} (2010), 217-222.

\bibitem{msr} C. Moreno and O. Sarbach, Stability properties of black holes in self-gravitating nonlinear electrodynamics, \textit{Phys. Rev.} D \textbf{67} (2003) 024028.

\bibitem{eir} E. F. Eiroa and C. M. Sendra, Gravitational lensing by a regular black hole, \textit{Class. Quant. Grav.} \textbf{28} (2011) 085008.

\bibitem{sha3} M. F. Shamir and G. Mustafa, Charged anisotropic Bardeen spheres admitting conformal motion,  \textit{Ann. Phys.} \textbf{418} (2020) 168184.

\bibitem{kum}  A. Kumar, et al., Bardeen black holes in the regularized $4D$ Einstein–Gauss–Bonnet gravity, \textit{MDPI.} D \textbf{8} (2022) 232.

\bibitem{ged} S. Gedela et al., Relativistic modeling of the neutron star in $Vela X-1$ via Bardeen space-time satisfying the embedding condition, \textit{New Astro.} (2021) 101583.

\bibitem{mus} G. Mustafa, M. F. Shamir, and M. Ahmad, A comparative analysis of self-consistent charged anisotropic spheres,  \textit{Phys. Dark. Uni.} \textbf{30} (2020) 100652.

\bibitem{sha2} M. F. Shamir and A. Malik, Bardeen compact stars in modified $f(R)$ gravity, \textit{Chi. J. Phys.} \textbf{69} (2021), 312-321.

\bibitem{carr} S. M. Carroll, V. Duvvuri, M. S. Trodenn and M. S. Turner, Is cosmic speed-up due to new gravitational physics? \textit{Phys. Rev.} D \textbf{70} (2004) 043528.

\bibitem{cap} S. Capozziello, Curvature quintessence, \textit{Int. J. Mod. Phys.} D \textbf{11} (2002), 483-491.

\bibitem{sha5} M. F. Shamir and I. Fayyaz, Analysis of compact stars in logarithmic-corrected $R^2$ gravity, \textit{Mod. Phys. Lett. A.} \textbf{35} (2019) 1950354.

\bibitem{nor} S. Nojiri and S. D. Odintsov, Modified gravity with negative and positive powers of curvature: Unification of inflation and cosmic acceleration, \textit{Phys. Rev.} D \textbf{68} (2003) 123512.

\bibitem{bert} O. Bertolami, C. G. Bohmer, T. Harko and F. S. N. Lobo, Extra force in $f(R)$ modified theories of gravity, \textit{Phys. Rev.} D \textbf{75} (2007) 104016.

\bibitem{odi} S. D. Odintsov. et al., Logarithmic-corrected $R^2$ gravity inflation in the presence of Kalb-Ramond fields, \textit{JCAP.} \textbf{02} (2019) 017.

\bibitem{ata} K. Atazadeh and F. Darabi, Energy conditions in $f(R,G)$ gravity, \textit{ Gen. Rel. and Gra.} \textbf{46} (2014), 1-14.

\bibitem{nor2} S. Nojiri and S. D. Odintsov, Modified Gauss–Bonnet theory as gravitational alternative for dark energy, \textit{Phys. Lett. B.}  \textbf{631} (2005), 1-6.

\bibitem{har} T. Harko, et al., $f(R,T)$ gravity, \textit{Phys.Rev.} D \textbf{84} (2011) 024020.

\bibitem{das} A. Das, et al., Gravastars in $f(R,T)$ gravity \textit{Phys. Rev.} D \textbf{95} (2017) 124011.

\bibitem{sha4} M. F. Shamir, $f(\mathcal {R},\varphi,\chi)$ cosmology with Noether symmetry, \textit{Eur. Phys. J. C.} \textbf{80} (2020) 1-9.
\bibitem{bun} H. A. Buchdahl, Non-linear Lagrangians and cosmological theory, \textit{Mon. Noti. R. Astron. Soc.} \textbf{150} (1970), 1--8.

\bibitem{noj} S. Nojiri and S. D. Odintsov, Unified cosmic history in modified gravity: from $F(R)$ theory to Lorentz non-invariant models, \textit{Phys. Rept.} \textbf{505} (2011), 59-144.

\bibitem{cog} G. Cognola et al., Class of viable modified $f(R)$ gravities describing inflation and the onset of accelerated expansion \textit{Phys. Rev.} D \textbf{77} (2008) 046009.

\bibitem{asta22} A. V. Astashenok, S. Capozziello, and S. D. Odintsov, Nonperturbative models of quark stars in $f(R)$ gravity,  \textit{Phy. Lett. B.} \textbf{742} (2015), 160--166.

\bibitem{asta44} A. V. Astashenok, S. Capozziello, and S. D. Odintsov, Further stable neutron star models from $f(R)$ gravity, \textit{J. Cosmo. and Astro. Phy.} \textbf{12} (2013) 040.

\bibitem{capo222} S. Capozziello, et al., The Mass-Radius relation for Neutron Stars in $f(R)$ gravity, \textit{Phy. Revi.} D \textbf{93} (2016) 023501.

\bibitem{asta11} A. V. Astashenok, et al., Maximum baryon masses for static neutron stars in $f(R)$ gravity, \textit{Eur. Lett.} \textbf{136} (2022) 59001.

\bibitem{asta33}  A. V. Astashenok et al., Causal limit of neutron star maximum mass in $f(R)$ gravity in view of GW190814, \textit{Phy. Lett. B.} \textbf{816} (2021) 136222.


\bibitem{herr} L. Herrera, J. Jimenez, L. Leal, J. Ponce de Leon, M. Esculpi and V. Galina, Anisotropic fluids and conformal motions in general relativity, \textit{J. Math. Phys.} \textbf{25} (1984), 3274--3278.

\bibitem{herr1}L. Herrera and J. Ponce de Leon, Anisotropic spheres admitting a one‐parameter group of conformal motions, \textit{J. Math. Phys.} \textbf{26} (1985), 2018-2023.

\bibitem{herr2} L. Herrera and J. Ponce de Leon, Isotropic and anisotropic charged spheres admitting a one‐parameter group of conformal motions, \textit{J. Math. Phys.} \textbf{26} (1985), 2302-2307.

\bibitem{Sha} M. F. Shamir, Massive compact Bardeen stars with conformal motion, \textit{Phys. Lett. B.} \textbf{811} (2020) 135927.

\bibitem{esc} M. Esculpi and E. Aloma, Conformal anisotropic relativistic charged fluid spheres with a linear equation of state, \textit{ Eur. Phys. J. Plus.} \textbf{67} (2010), 521-532.

\bibitem{mak} M. K. Mak and T. Harko, Quark stars admitting a one-parameter group of conformal motions, \textit{Int. J. Mod. Phys.} \textbf{13} (2004), 149-156.


\bibitem{chnd} S. Chandrasekhar, Erratum: the Dynamical Instability of Gaseous Masses Approaching the Schwarzschild Limit in General Relativity, \textit{Astrophys. J.} \textbf{140} (1964) 1342.

\bibitem{buchd} H. A. Buchdahl, General relativistic fluid spheres, \textit{Phy. Rev.} \textbf{116} (1959) 1027.

\bibitem{makh11} M. K. Mak and T. Harko, Anisotropic stars in general relativity, \textit{Proc. R. Soc. Lond.} \textbf{459} (2003), 393-408.

\bibitem{smh} S. M. Hossein, et al., Anisotropic compact stars with variable cosmological constant, \textit{Int. J. Mod. Phys.} \textbf{21} (2012) 1250088.

\bibitem{star2} A. A. Starobinsky, A new type of isotropic cosmological models without singularity, \textit{Phys. Lett. B.} \textbf{91} (1980), 99-102.








\bibitem{corm} V. F. Cormier, M. I. Bergman and P. L. Olson, Earth's Core, \textit{Elsevier} (2022) 1.




\bibitem{and} C. Andreasson, Sharp bounds on the critical stability radius for relativistic charged spheres, \textit{ Math. Phys.} \textbf{288}  (2009), 715-730.

\bibitem{iva} B. V. Ivanov, Static charged perfect fluid spheres in general relativity, \textit{Phys. Rev. D.} \textbf{65}  (2002) 104011.



\end{thebibliography}
\end{document}